\documentclass[twocolumn]{aastex701}

\usepackage{xcolor}
\usepackage{amsmath}
\usepackage{array}
\usepackage{stackengine}
\usepackage{booktabs}
\usepackage{float}

\begin{document}

\title{The Gamma-ray Burst Jet Energy Distribution Suggests A Quasi-universal, Weakly Magnetized Jet Evolving Over Cosmic Time}


\author[orcid=0000-0003-1707-7998]{Nicole Lloyd-Ronning}
\affiliation{Computational Physics and Methods Group, Los Alamos National Lab, Los Alamos, NM 87544}
\affiliation{Center for Theoretical Astrophysics, Los Alamos National Lab, Los Alamos, NM 87544}
\email[show]{lloyd-ronning@lanl.gov}

\author[0000-0002-3137-4633]{Fabio De Colle} 
\affiliation{Instituto de Ciencias Nucleares, Universidad Nacional Autonoma de Mexico, Apartado Postal 70-543, Ciudad de Mexico 04510, Mexico}
\email{fabio@nucleares.unam.mx}

\author[orcid=0009-0001-0792-8341]{Gal Birenbaum}
\affiliation{Astrophysics Research Center of the Open University (ARCO), The Open University of Israel, P.O. Box 808, Ra’anana 4353701, Israel}
\affiliation{Department of Natural Sciences, The Open University of Israel, P.O. Box 808, Ra’anana 4353701, Israel}
\email{birenbaumgal@gmail.com}

\author[orcid=0000-0003-4271-3941]{Omer Bromberg}
\affiliation{The Raymond and Beverly Sackler School of Physics and Astronomy, Tel Aviv University, Tel Aviv 69978, Israel}
\email{omerbr@tauex.tau.ac.il} 

\author[0000-0001-8002-2661]{Jarrett Johnson}
\affiliation{XTD Nuclear Threat Assessment, Los Alamos National Lab, Los Alamos, NM 87544}
\affiliation{Center for Theoretical Astrophysics, Los Alamos National Lab, Los Alamos, NM 87544}
\email{jlj@lanl.gov}


\begin{abstract}

We present distributions of gamma-ray burst observed and inferred properties for those GRBs with redshifts.  We show that the isotropic energy distribution, which spans over four orders of magnitude, can be reproduced reasonably well under a simplistic assumption that {\em every observed GRB originates from a quasi-universal jet} with roughly the same decreasing power-law profile of energy as a function of angle, with a power-law index of $ 3 \lesssim  \zeta \lesssim 4$. The spread in the observed distribution can be explained by the variation in observer viewing angle alone. Furthermore, this power-law jet structure provides an even better fit to both the isotropic energy and luminosity distributions if the isotropic energy normalization {\em evolves as a function of redshift} in a manner that has been suggested by a range of previously published studies.  The relatively steep power-law index of this jet is consistent with the structure predicted by simulations of {\em weakly magnetized} jets in collapsars, whereas simulations of hydrodynamic jets predict a structure shallower than what we find here. The predicted afterglow light curves within this model framework show steepening behavior at times commensurate with observed jet break times.

\end{abstract}



\section{Introduction} 

Gamma-ray bursts, the most luminous objects in our universe, provide insight into the extreme physics of relativistic outflows and the end states of massive stars.  Detailed analysis of thousands of these transient events suggests that the highly variable prompt gamma-ray emission (in the $\sim 100$'s of keV range, lasting $\sim 10$'s of seconds) arises from internal dissipation processes in a relativistic jet, while the long lived afterglow (spanning the entire electromagnetic spectrum, fading over months to years) originates from the decelerating shock system (forward and reverse shock) at the head of the jet.  For relatively comprehensive reviews of GRBs, see \cite{Pir04,Mesz06, WB06, GRRF09, Berg14, KZ15, Lev16}.  

Despite this generally well-supported paradigm, there exists a number of unsolved questions regarding the nature of their progenitor systems, their central engines, and the underlying microphysics in their outflows. Studies have directly associated some GRBs to massive stars due to the presence of supernovae accompanying GRB events \citep{Hjorth03, WB06,HB12}, and/or other telltale massive star signatures like a plateau in the prompt duration distribution \citep{Brom12, Brom13, LRBP26}. Other GRBs have been convincingly connected to double neutron star mergers or neutron star-black hole mergers due to the coincident detection of gravitational waves \citep{Ab17}, or signatures of heavy-element production expected in extremely neutron-rich compact object merger environments \citep[e.g.][although see \citealt{Rist26} who show that two long GRBs with kilonovae signatures can be explained in the context of a collapsar model]{Tro19, OC21, Rast22}. Many studies have also examined correlations among GRB variables, including how these variables evolve over cosmic time, with the intent to gain a different perspective or insight into the physics behind these events \citep[for a relatively comprehensive review of GRB correlations, see][]{Dai19}. \\

However, in general, GRBs do not lend themselves cleanly to progenitor categorization, related to the fact that a variety of systems are (at least theoretically) capable of creating a central engine that can produce a GRB-like jet or event.  There appears to be a lack of uniformity (i.e. no ``standard candle'') in any particular GRB property or behavior, including duration, variability, time-resolved and time-averaged spectral behavior, overall energy and luminosity, etc.\footnote{i.e. If you've seen one, you've seen...one.}  An intriguing counter-example was reported in the early 2000's, when only a handful of GRBs had redshift measurements. 
Several studies \citep{Frail2000, Piran01, BKF03, BlFK03} showed that GRB beaming-corrected jet energy seemed to cluster narrowly around a value of about $5 \times 10^{50}$ erg.  Subsequent studies, with the advantage of more data, new measurement techniques, and additional consideration of selection effects, suggested that the distribution of GRB jet energy is in fact much broader \citep{Ghir04, Nava06, Li08, Racusin09, Shiv11, Wang18}. Nonetheless, all of these works opened up the interesting problem of determining the physics behind the shape, width, and peak of the GRB jet energy distribution, and how this may connect to the central engine and progenitor system.\\

 We aim to at least partially address that question in this paper. We first present the observed (or inferred) distributions of both isotropic-equivalent and beaming-corrected jet energy and luminosity, as well as inferred GRB jet opening angles. We then test whether the isotropic-equivalent energy distribution can be explained by a universal GRB jet, where the only variable is observer viewing angle. We show that this distribution is fairly well reproduced by a smoothly broken power-law jet structure (energy as a function of angle from the jet axis). This same structure fails to accurately capture the isotropic-equivalent peak luminosity distribution.  However, when we include evolution of the normalization  with redshift, according to the functional form of observed evolution reported in a number of published studies, both the isotropic energy and luminosity distributions are well-reproduced by a ``universal'' power-law jet structure. We discuss how this structure might be physically meaningful, drawing on previous high fidelity simulations of both hydrodynamic and magnetized jets.  We also explore the implications for the predicted afterglow from such jets, and present some representative light curves based on this jet model in optical and X-ray frequencies.  We show that the predicted jet break times in the light curves are roughly commensurate with the distribution of observed GRB jet break times, particularly if we relax our overly stringent assumptions of a single set of microphysical parameters and/or external circumburst density.\\

This paper is organized as follows: In \S 2, we describe our data sample and definitions of the measured and inferred variables. We present the distributions of isotropic and beaming-corrected energy and luminosity, as well as the inferred jet opening angles (calculated from breaks in the afterglow light curves under a uniform ``top-hat" jet assumption).  In \S 3, we present the predicted isotropic equivalent energy and peak luminosity distributions for a universal power-law jet, both with and without redshift evolution, and show how it compares to the observed data.  In \S 4, we present representative afterglow light curves from this universal jet, and generate a distribution of predicted break times, which we compare to observed break times. In \S 5, we provide a discussion of the physical implications of our results, comparing to both observations of jet structure from individual GRBs as well as predictions from high fidelity simulations.  Finally, in \S 6, we present a summary and our conclusions.

\section{Data and Distributions}
Our data are taken from \cite{Wang20}, who have compiled publicly available observations of $6289$ gamma-ray bursts from 1991 to 2016.   We searched this dataset for those GRBs with measured redshifts, the majority of which are GRBs observed by the {\em Neil Gehrels Swift} Observatory (where there are 492 {\em Swift} GRBs with redshifts out of the full sample of 567 GRBs with redshift values).

\subsection{Energy and Luminosity Definitions}
The isotropic-equivalent energy $E_{\rm iso}$ is calculated from the measured prompt gamma-ray fluence $F$ (i.e. time integrated flux) and its redshift, assuming the emission is isotropic over $4\pi$ steradians:\\
\begin{equation}
    F = \frac{\xi E_{\rm iso}(1+z)}{4 \pi d_{L}^{2}}\;, 
\end{equation}
\noindent where $\xi$ is an efficiency factor quantifying how much of the jet's total energy goes into radiating the gamma-ray emission, and the  factor of $(1+z)$ comes from the time integral of flux, accounting for cosmological time dilation. The luminosity distance $d_{L}$ is given by
\begin{equation}
    d_L = (1+z) \frac{c}{H_0} \int_0^z \frac{dz'}{ \sqrt{\Omega_M(1+z')^3 + \Omega_\Lambda}}
\end{equation}
\noindent where $c$ is the speed of light, $H_{0}$ is the Hubble constant, $\Omega_{m}$ and $\Omega_{\Lambda}$ are the fraction of energy in matter and dark energy respectively.  We adopt a standard cosmology with $H_{0} = 67.3$ km s$^{-1}$ Mpc$^{-1}$, $\Omega_m = 0.315$, and $\Omega_{\Lambda} = 0.685$. \\

\begin{figure}
    \includegraphics[width=0.47\textwidth]{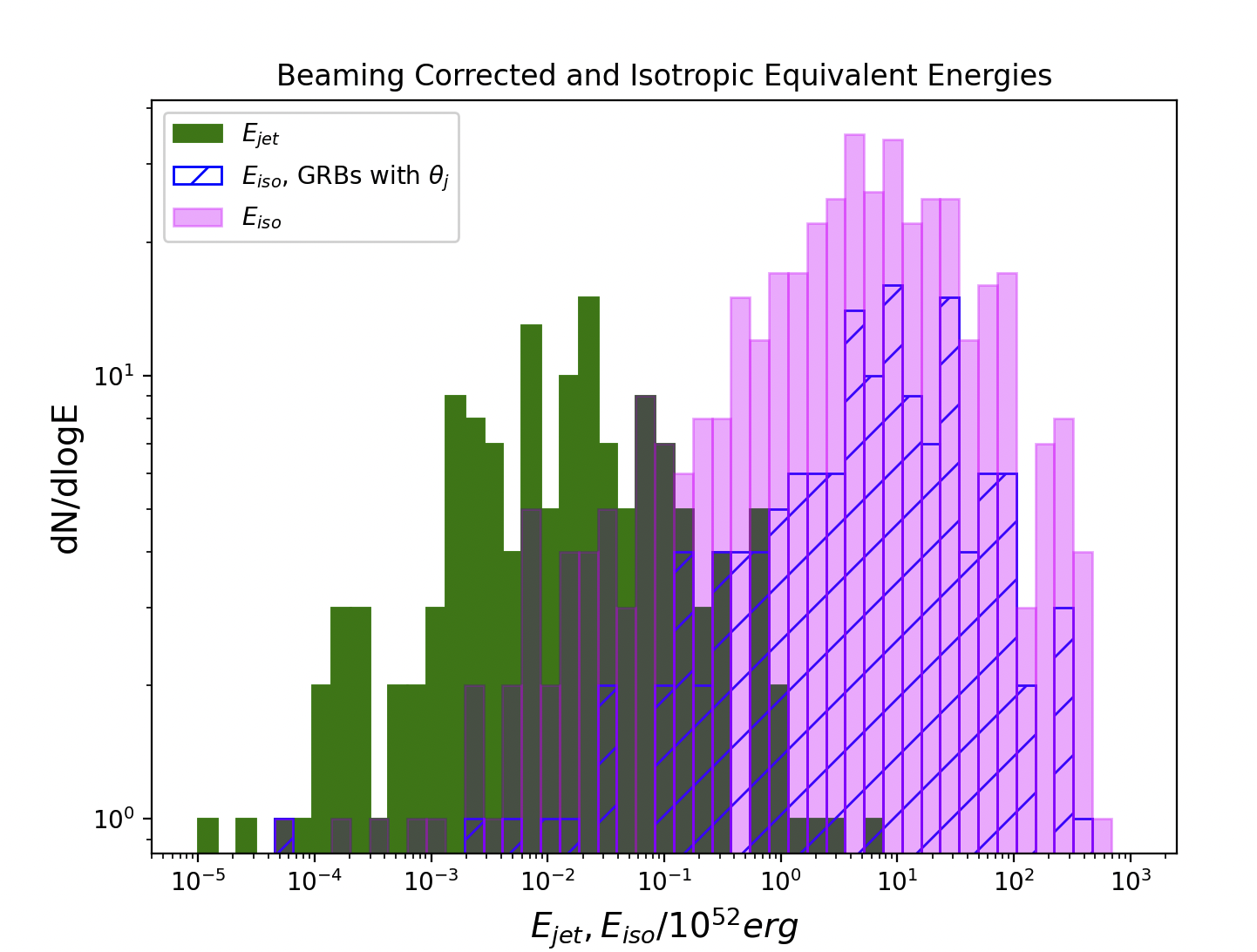}
    \caption{Distributions of $E_{\rm iso}$ and beaming-corrected energy $E_{\rm jet}$. The pink histogram shows the isotropic equivalent energy, while the blue hatched histogram shows this distribution for the subset of GRBs with jet opening angle measurements.  The green histogram in this figure is the beaming-corrected jet energy.}
    \label{fig:EisoEbeam}
\end{figure}

\begin{figure}
    \includegraphics[width=0.45\textwidth]{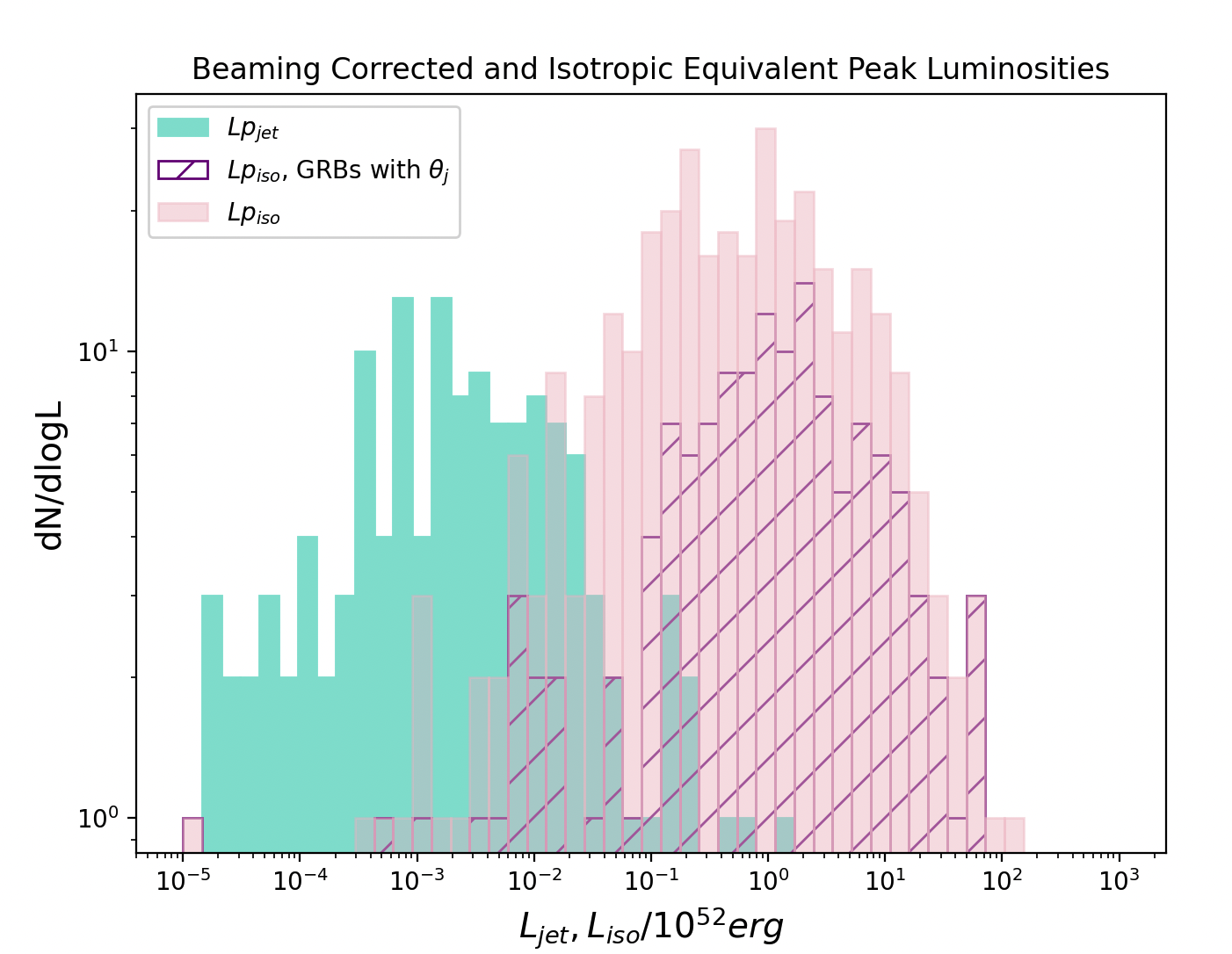}
    \caption{Distributions of $L_{\rm iso}$ and beaming-corrected luminosity $L_{\rm jet}$. The light pink histogram shows the isotropic equivalent luminosity, while the purple hatched histogram shows this distribution for the subset of GRBs with jet opening angle measurements.  The cyan histogram in this figure is the beaming-corrected jet luminosity.}
    \label{fig:LisoLbeam}
\end{figure}

\begin{figure}
    \includegraphics[width=0.47\textwidth]{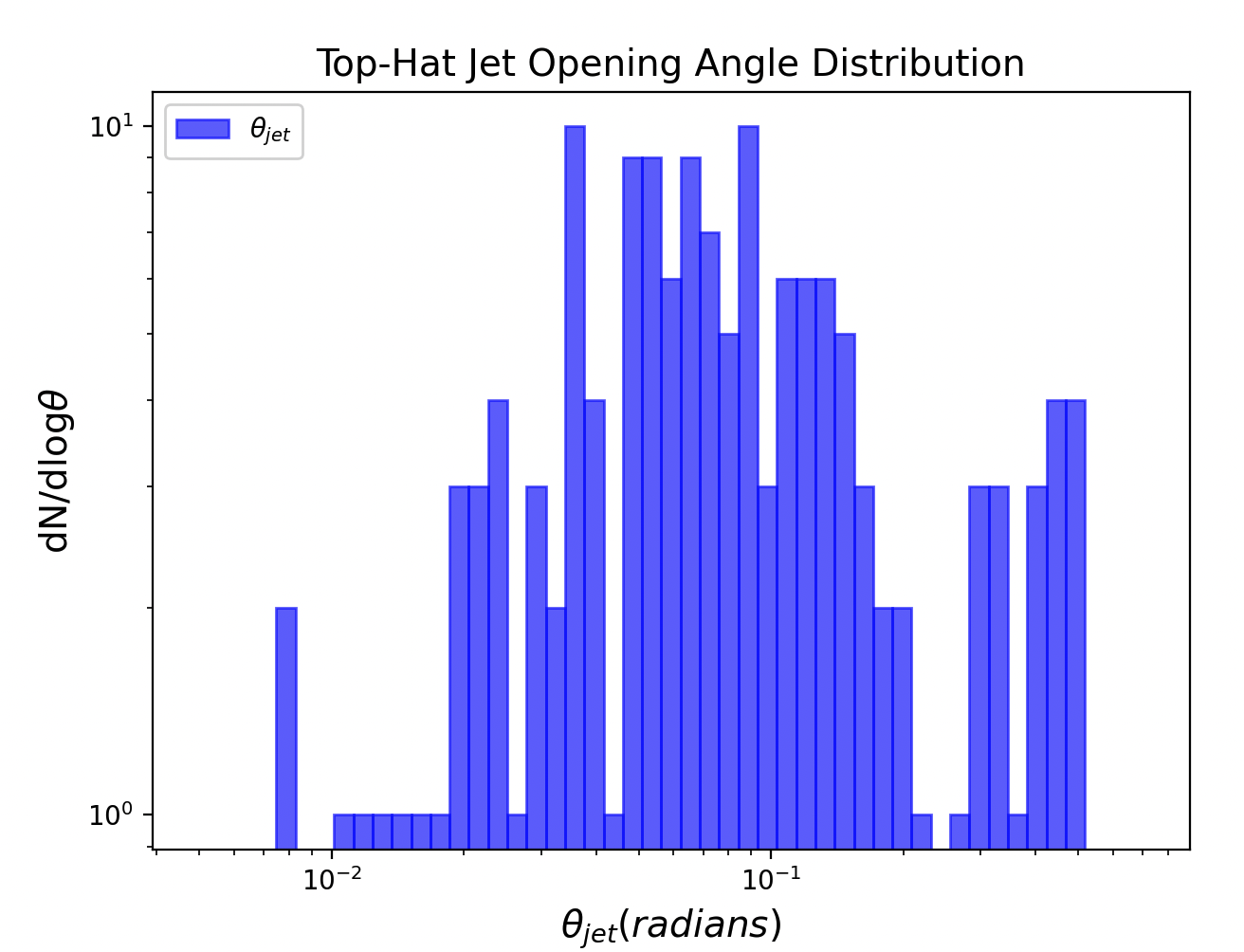}
    \caption{Distributions of jet opening angle measurements, inferred within a top-hat uniform jet model as described in section 2.2.}
    \label{fig:thetaj}
\end{figure}

Similarly the isotropic-equivalent peak luminosity $L_{p,\rm iso}$ is computed through the relationship:\\
\begin{equation}
    f_{p}  = \frac{\xi  L_{p, \rm iso}}{4 \pi d_{L}^{2}}
\end{equation}
\noindent where $f_{p}$ is the peak flux during the prompt gamma-ray burst phase.\\

The beaming-corrected energy and luminosity, $E_{\rm jet}$ and $L_{\rm jet}$, are then given by:
\begin{equation}
    [E_{\rm jet}, L_{p,\rm jet}] = (1-\cos\theta_{j})[E_{\rm iso}, L_{p,\rm iso}]\;,
\end{equation}
\noindent where $\theta_{j}$ is the inferred half-opening angle of the jet (discussed in the following subsection). \\

Figures~\ref{fig:EisoEbeam} and ~\ref{fig:LisoLbeam} show the distributions of energy and peak luminosity.  In Figure~\ref{fig:EisoEbeam}, the pink histogram shows the isotropic equivalent energy, while the blue hatched histogram shows this distribution for the subset of GRBs with jet opening angle measurements.  The green histogram in this figure is the beaming-corrected jet energy.  A similar scheme is shown for the luminosity distributions in Figure~\ref{fig:LisoLbeam}. \\

The energy and luminosity distributions are relatively wide. The isotropic-equivalent energy distribution peaks at an energy close to  $10^{53}$ ergs, while the beaming-corrected jet energy peaks at about $2 \times 10^{50}$ ergs; both have a spread of over 4 orders of magnitude.  The isotropic-equivalent jet peak luminosity peaks close to $10^{52}$ erg s$^{-1}$, while the beaming-corrected jet luminosity peaks at around $5 \times 10^{49}$ erg s$^{-1}$; the luminosity distributions are a bit narrower than the energy distributions, spread over about 3 to 3.5 orders of magnitude.

\subsection{Jet Opening Angle}
The jet opening angle ($\theta_{j}$) values in the literature are typically estimated from a break or steepening observed in the afterglow light curve. Most studies have calculated this value under the assumption that the jet has a top-hat structure, with constant energy across the jet up to a given angle and a step-function drop off beyond that angle.  Then, the time of the break reflects when the forward blast wave has decelerated to a point at which the relativistic beaming of the radiation is comparable to the physical opening angle of the jet. In other words, when the outflow velocity of the jet has decelerated to a point where relativistic beaming ($\sim 1/\Gamma$, where $\Gamma$ is the jet bulk Lorentz factor) is on the order of the physical jet opening angle $\theta_{j}$, photons are able to escape “sideways” and a steepening of the light curve will occur \citep{Rhoads97,Rhoads1999}. 

The jet opening angle in this top-hat jet framework can be calculated from:
\begin{multline}
    \theta_j = 0.069 \left(\frac{t_{\rm b}}{1 \text{ day}}\right)^{3/8} 
    \left(\frac{1+z}{2}\right)^{-3/8} 
    \left(\frac{E_{\text{iso}}}{10^{52} \text{ erg}}\right)^{-1/8} \\
    \left(\frac{\xi}{0.01}\right)^{1/8} 
   \left(\frac{n}{1 \text{ cm}^{-3}}\right)^{1/8} \text{ rad}
\end{multline}
\noindent where $t_{b}$ is the timescale of the steepening or break in the light curve and $n$ is the density of the medium. \\

Figure~\ref{fig:thetaj} shows the distributions of jet opening angles calculated from afterglow light curve break times.  The distribution peaks around 0.1 radians ($5.7^{\rm o}$), and is spread relatively broadly over 2 orders of magnitude with an average value of 0.12 radians ($6.8^{\rm o}$).   

\begin{table*}[t]
\centering
\caption{Power-law (PL) models (equation 6) for the GRB isotropic equivalent energy distribution. The ``z Distribution'' column indicates whether we used the full redshift (z) distribution available, whether we split our sample into low and higher redshift subgroups. The Cosmic evolution column indicates how we incorporated cosmological evolution of either the normalization factor or characteristic jet angles, according to the dependence  suggested in the literature. These are the representative jet structure models we use to produce the hatched histograms in Figures 4-6 of the text; see the exclusion analysis in the Appendix for a more detailed account of the error region and model degeneracies.}
\begin{tabular}{lccccccc}
\hline
Model & z Distribution & $A (10^{55}$ erg) & Efficiency $\xi$ &$\theta_{\rm m}$ (radians) & $\theta_{j}$ (radians) & {\bf PL Index $\zeta$} & Cosmic evolution \\
\hline \hline \hline
PL 1 & full & $2.45 $ & $0.01$ &  $0.021 $ & $0.41  $ & $3.53 \pm 0.5 $ & No\\
PL 2 & full & $1.00 $ & $0.005$ &  $0.021 $ & $0.41  $ & $3.00 \pm 0.5$ & No \\
PL 3 &  $ z > 1.5$ & $4.056 $ & $0.01$ &  $0.018 $ & $0.414  $ & $3.52 \pm 0.50$ & No \\
PL 4 &  $z < 1.5$ & $1.056 $ & $0.01$ &  $0.018 $ & $0.414  $ & $3.52 \pm 0.50$  & No\\
PL 5 & full & $0.50 $ & $0.005$ &  $0.020 $ & $0.41 $ & $3.50 \pm 0.50$ & $A \propto (1+z)^{1.5}$\\
PL 6 & full & $0.40 $ & $0.01$ &  $0.020 $ & $0.41 $ & $3.50 \pm 0.50$ & $\theta_{*} \propto (1+z)^{-1.0}$\\
%
%
\hline \hline
\end{tabular}
\end{table*}

\begin{table*}[ht!]
\centering
    \caption{Representative power law models for the GRB isotropic equivalent peak luminosity distribution, used to produce the hatched histograms shown in Figure 7.}
    \begin{tabular}{lccccccc}
        \hline
        Model & z Distribution & $A (10^{53}$ erg) & Efficiency $\xi$ &$\theta_{\rm m}$ (radians) & $\theta_{j}$ (radians) & PL Index $\zeta$ & $L_{\rm p,iso}$-z Evolution \\
        \hline
        \hline \hline
        PL A & full & $50. $ & $0.01$ &  $0.021 $ & $0.41 $ & $3.5 \pm 1.50$ & No\\
        PL B & full & $.33 $ & $0.005$ &  $0.010 $ & $0.21  $ & $2.5 \pm 1.00$ & $(1+z)^{3.0}$ \\ \hline \hline
    \end{tabular}
\end{table*}

\section{Can We Reproduce the Observed Distributions from a ``Universal'' Structured Jet?}

We would like to test whether a ``universal'' structured jet model, where every GRB has the same energy distribution as a function of angle from the jet core axis, is capable of reproducing our observed energy and luminosity distributions. In this case, the spread in the distributions would simply be a result of the variation in observer viewing angle.  In the analysis below, we use only the Swift GRBs, so that we can apply a uniform detector flux and/or fluence limit when deciding if a simulated GRB will trigger the detector.
The following describes the approach we use to generate a mock observed isotropic-equivalent energy distribution from a structured jet:

\begin{figure*}
\begin{centering}
    \stackinset{r}{6.9cm}{t}{2.1cm}{   \includegraphics[width=0.09\textwidth]{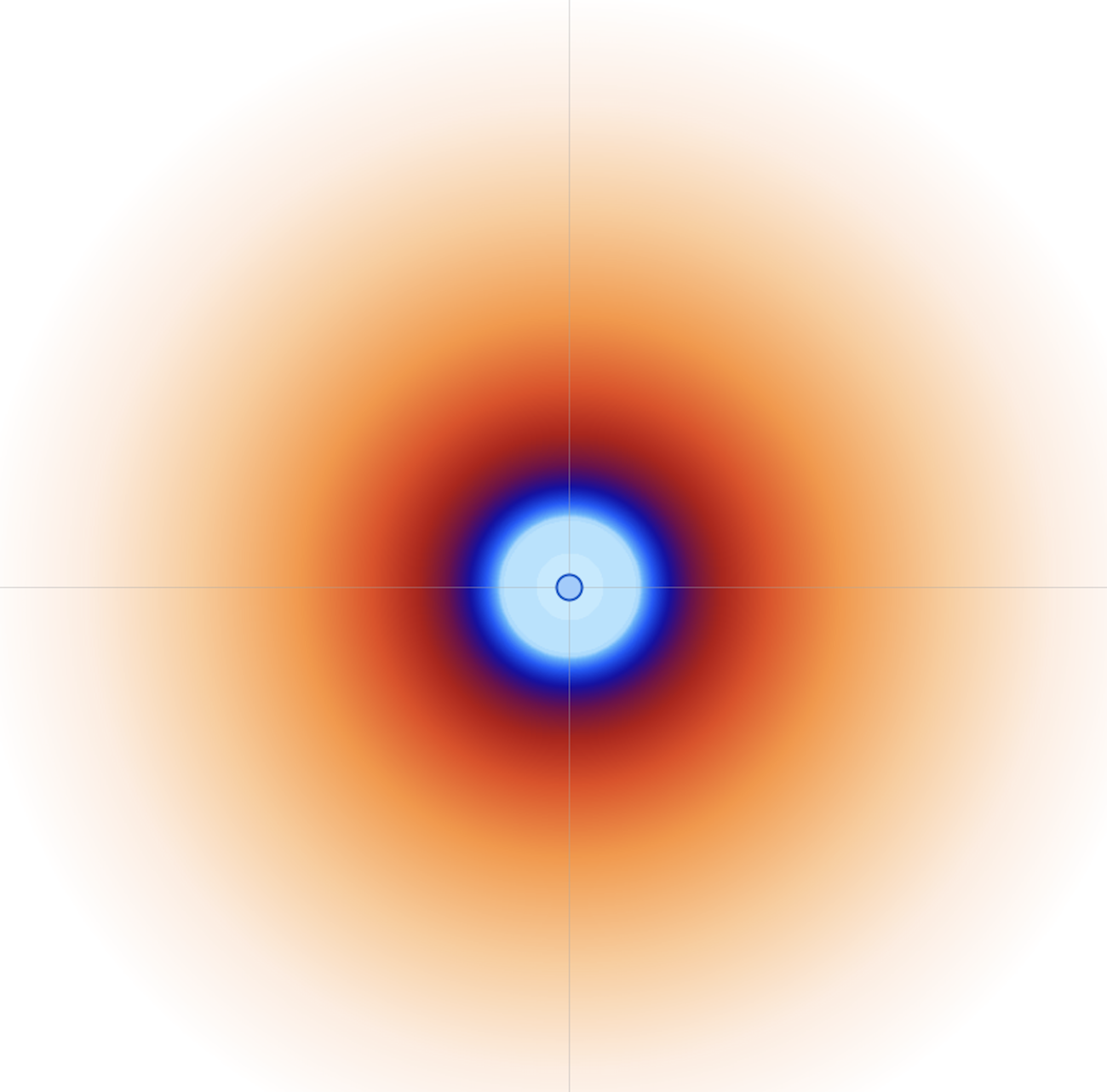}
  } {\stackinset{r}{4.9cm}{t}{3.2cm}{   \includegraphics[width=0.19\textwidth]{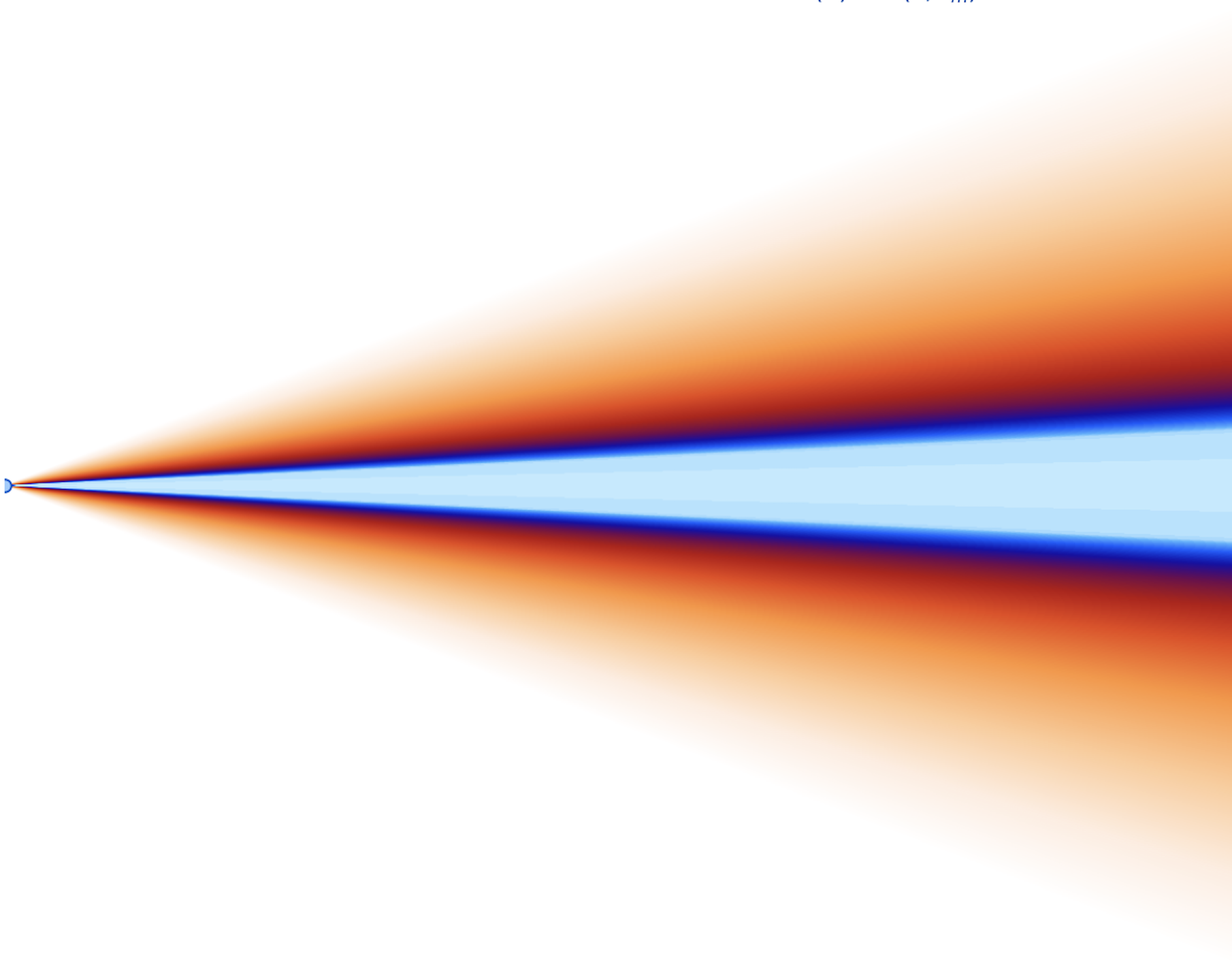}
  }
  {  \includegraphics[width=0.6\textwidth, height=7.5cm]{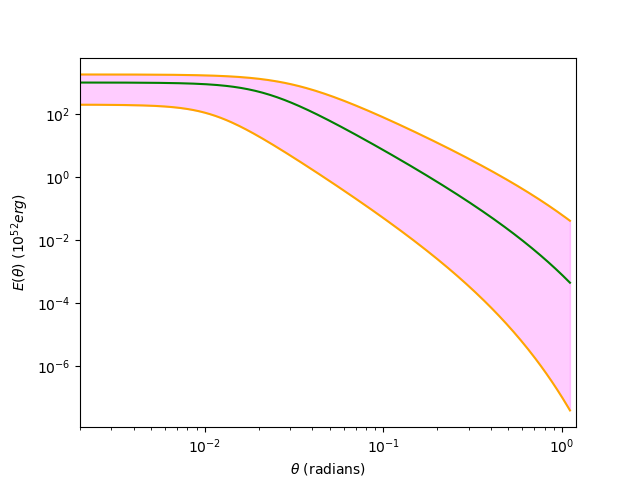}}}\\
  \includegraphics[width=0.47\textwidth]{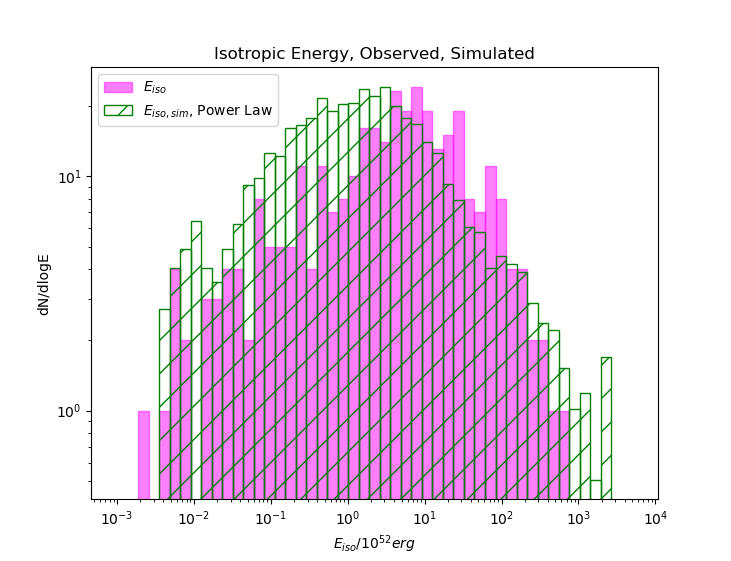}
 \includegraphics[width=0.47\textwidth, height=6.5cm]{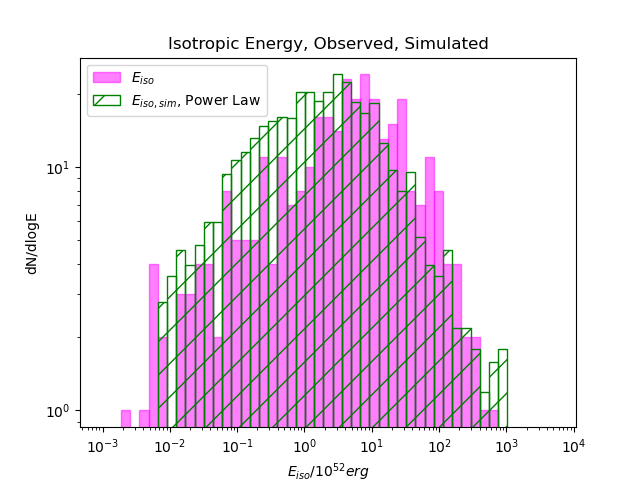}
    \caption{
    {\bf Upper:} Functional form of the energy distribution with angle from the central jet axis, shown by the green line, with the error bar region shown by the magenta shading.  The inset shows a schematic of the face-on and side view of this jet structure.  {\bf Lower:} The resulting predicted isotropic-equivalent energy distribution (blue hatched histogram) compared to the observed distribution (magenta histogram), for Model PL 1 (left) and Model PL 2 (right).  Both require relatively steep power-law indices for the jet energy profile,  $3 \lesssim \zeta \lesssim 4$. We note the redshift distribution of the surviving simulated sample is commensurate with the observed/parent redshift distribution.}
    \label{fig:jetstruct}
\end{centering}
\end{figure*}

\begin{figure*}
\centering
    \includegraphics[width=0.49\textwidth]{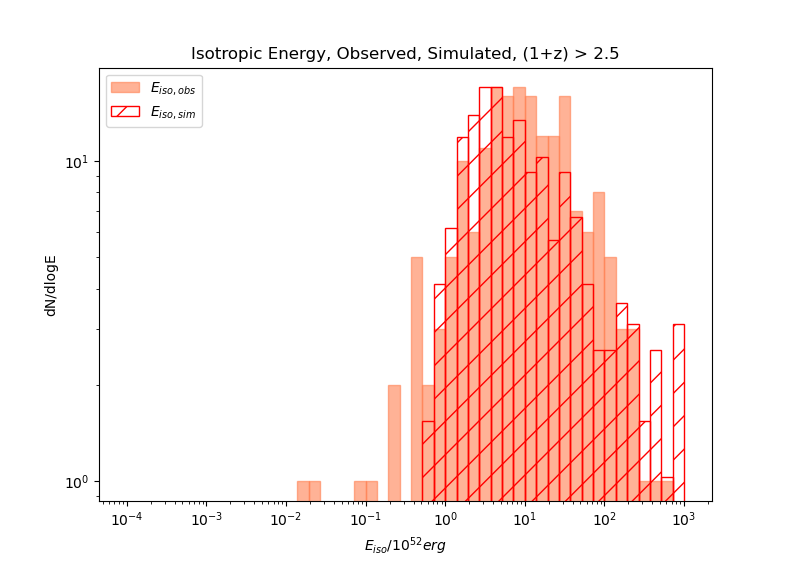}\includegraphics[width=0.47\textwidth]{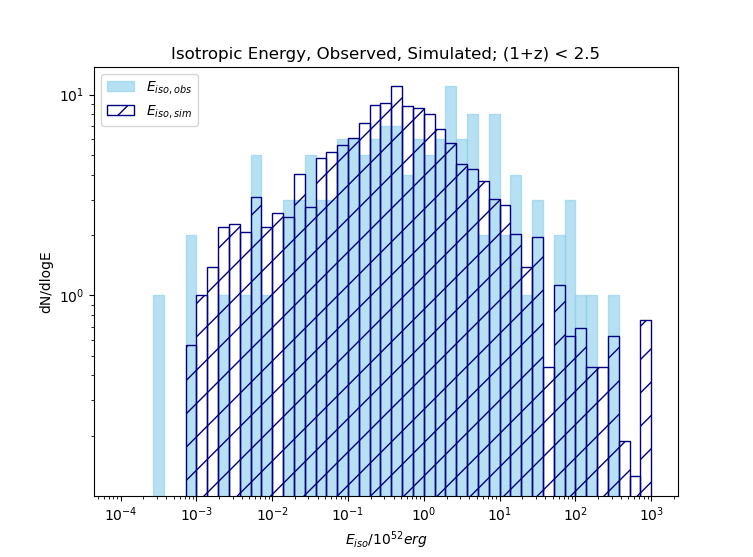}
    \caption{Higher redshift (left) and lower redshift (right) sample, fit to the same power-law structure, but with different normalizations, described by models 3 and 4 in Table 1}
    \label{fig:lowzhiz}
\end{figure*}

\begin{figure*}
\centering
    \includegraphics[width=0.49\textwidth]{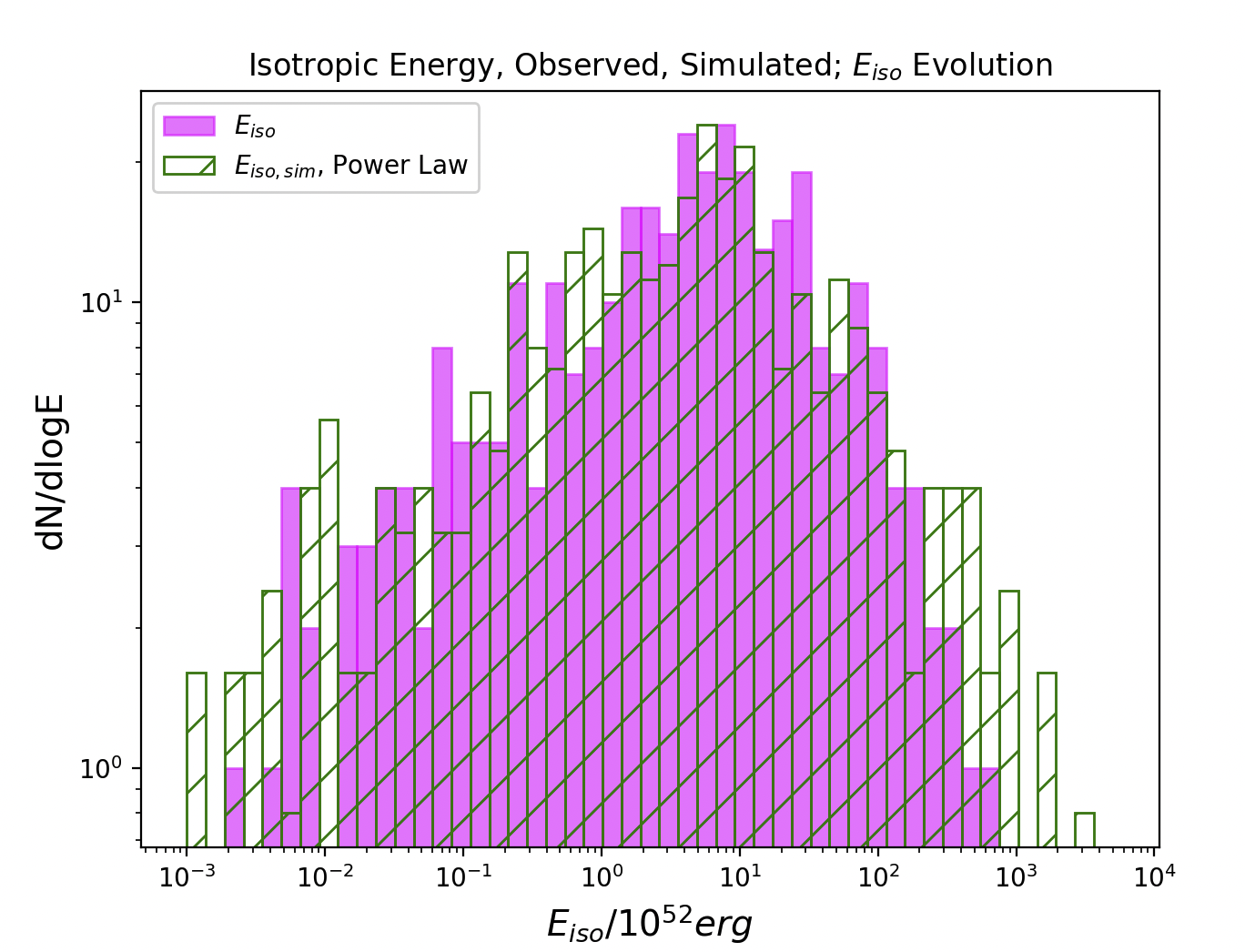}\includegraphics[width=0.49\textwidth,height=6.8cm]{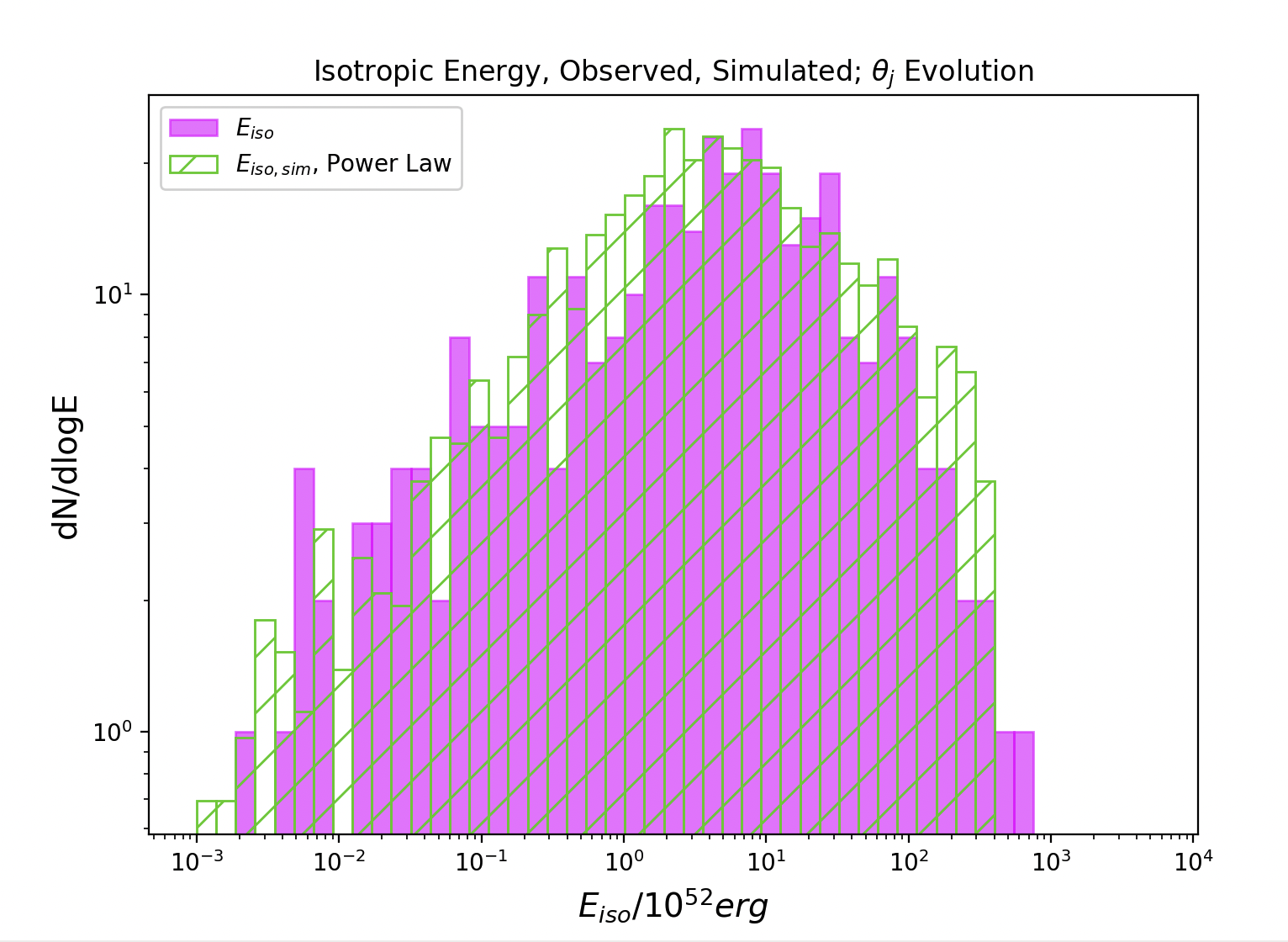}
    \caption{{\bf Left:} Simulated and observed $E_{\rm iso}$ distributions for a power-law jet structure but with a normalization that evolves with redshift according to previously published literature, $E_{\rm iso} \propto (1+z)^{1.5}$, according to Model PL 5 in Table 1. {\bf Right:} Simulated and observed $E_{\rm iso}$ distributions for a power-law jet structure but with the characteristic jet angles that evolve with redshift according to previously published literature, $\theta_{m,j} \propto (1+z)^{-1}$, described by Model PL 6 in Table 1.}
    \label{fig:eisoevol}
\end{figure*}

\begin{enumerate}
    \item Choose a  model for the jet structure.  We tried different simple, standardly-used functions like a Gaussian jet ($E(\theta) = A e^{-\frac{1}{2}(\theta/\theta_{j})^{2}}$), a simple power-law ($E(\theta) \propto \theta^{-\zeta}, \  \theta > \theta_{m}$; $E(\theta) = {\rm constant},\theta < \theta_{m}$ ), 
    and a power-law with an exponential cutoff.  The Gaussian models do not converge and cannot reproduce the width of the energy distribution, so we do not consider this model further.  The functional form of the jet structure that reproduces the energy distribution best (in terms of a Kolmogorov-Smirnov test between the observed and simulated distributions giving high probability that the two distributions are from the same parent distribution) is a power-law with an exponential cutoff:

\begin{equation}
E(\theta) = \frac{\text{A}}{1 + \left(\frac{\theta}{\theta_{m}}\right)^{\zeta}} \exp\left(-\frac{\theta}{\theta_{j}}\right)
\end{equation}

\noindent This function describes a jet that has a relatively constant core energy from the central axis out to an angle of $\theta_{m}$, and then decreases as a power law with an index $-\zeta$ beyond $\theta_{m}$ and, finally, has an exponential cutoff at a characteristic angle $\theta_{j}$.  The functional form and a visualization of this structure are shown in Figure~\ref{fig:jetstruct}.

    \item Pull the observer viewing angle $\theta_{v}$ from a uniform distribution (on a sphere), where the $\theta_{v}$ is the viewing angle from the central axis of the jet. This is justified because GRBs (and their jet orientations) are distributed isotropically across the sky and there is no preferred (cosmic) direction for the orientation of their jets on the sky.    
    
    \item Assume that the velocity structure of the jet is such that it is relativistic at every angle/line of sight, so that $1/\Gamma \ll \theta_{j}$ and the viewer is receiving the (prompt) radiation/energy only from this narrow cone over which the energy can be approximated as a constant,  $E(\theta_{v})$.  This assumption is supported by the work of \cite{Kat19, BN19, Kat24}.  In particular, Figure 2 of the latter reference shows $\Gamma > 50$ across the jet, even as the energy decreases as a function of angle.  We revisit the validity and implications of this assumption in \S 5.2 in the Discussion.

    \item Assign the GRB a redshift, according to the observed GRB redshift distribution, corrected for Malmquist bias as in \cite{LR19}\footnote{Their non-parametric technique to correct for truncation in the luminosity-redshift plane, using the Lynden-Bell \citep{LB71} and Efron-Petrosian \citep{ep98} methods, produces a slight correction to the GRB $dN/dz$ distribution at low redshifts. See their Figure 6.}.  We also, for comparison, assigned our GRBs a redshift under the assumption that the GRB rate density exactly traces the Madau-Dickinson star formation rate \citep{MD14}. Plots of both the observed GRB redshift distribution and the redshift distribution following the star formation rate (along with a discussion of the issues and biases in using each) are given in the Appendix.  We find that when using a GRB redshift distribution that traces the star formation rate, our jet model cannot reproduce the full width of the $E_{\rm iso}$ distribution.

    \begin{figure*}[ht!]
\centering
\includegraphics[width=0.46\textwidth, height=6.35cm]{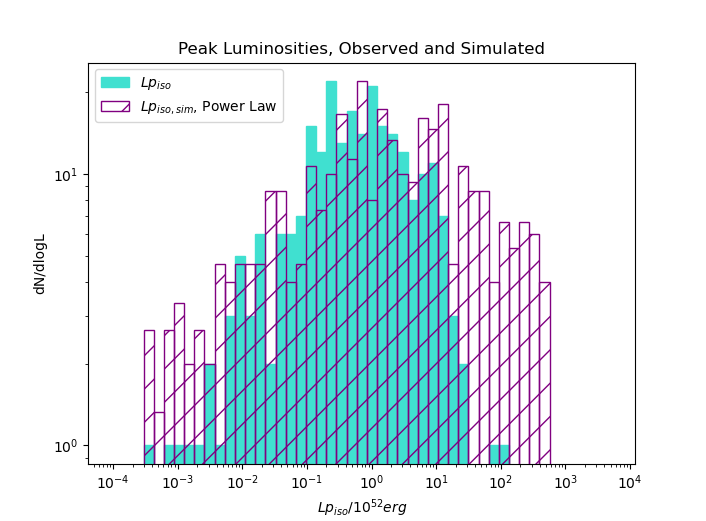}\includegraphics[width=0.45\textwidth]{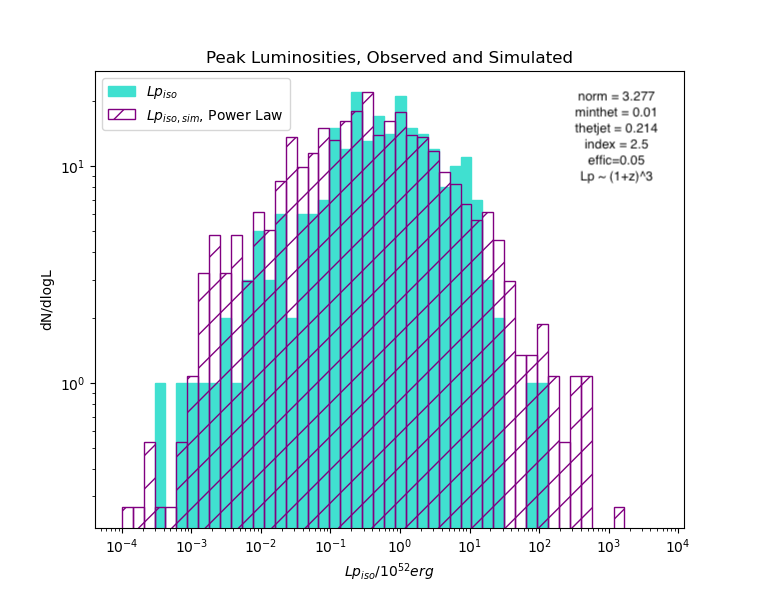}
\caption{{\bf Left:} The simulated isotropic-equivalent peak luminosity distribution (purple hashed histogram) compared to the observed distribution (cyan histogram) for the jet structure described by Model A in Table 3 (which is the same jet structure as Model 1 but with a reduced normalization). {\bf Right:} Model B in Table 3, but including evolution of the luminosity over redshift according to functional forms in the published literature, $L_{p,\rm iso} \propto (1+z)^{3.0}$.}
\label{fig:lum}
\end{figure*}

    \item For any given simulated energy distribution, assume a constant radiative efficiency factor across all GRBs. 
    We use either $\xi = 0.01$ or $\xi = 0.005$, as indicated in Table 1.  This is of course an overly simplistic assumption; however, we emphasize again that we are not trying to make a physical claim about the universality of the efficiency but rather test how well this type of single-parameter basic model does in reproducing the observed distributions we examine here.

    \item Given the energy (calculated from our viewing angle), redshift, and efficiency, calculate the GRB fluence according to equation 1.  If the fluence is above the nominal {\em Swift} BAT fluence limit of $\sim 10^{-8} {\rm erg \ cm^{-2}}$ store the energy in an array.\footnote{When we analyze the peak luminosity distribution, we trigger on the nominal flux limit of $\sim 2 \times 10^{-8} {\rm erg \ cm^{-2}}$.  We note that the Swift BAT flux and fluence limits are in reality not constants.  However, these values are generally accepted as reasonable approximations of a nominal detector fluence and/or flux limit.}

    \item Do this $n \sim 80000$ times. We iterate over the distribution of parameters using a Markov Chain Monte Carlo (MCMC) approach comparing each generated distribution of isotropic-equivalent energies to the observed distribution.  We use both a KS-test to compare the observed and simulated cumulative distributions, maximizing the p-value between the two distributions (where the p-value indicates the probability that the two distributions were drawn from the same parent distribution), and a typical maximum likelihood approach in our MCMC analysis.   
 
\end{enumerate}

The peak and spread in the $E_{\rm iso}$ distribution can be reasonably well reproduced by this very basic power-law model.  PL 1 and PL 2 in Table 1 show our two representative (non-evolving) models, with the histograms of the observed (pink) and simulated (green hatched) distributions shown in Figure~\ref{fig:jetstruct}.  Each of these models produces $E_{\rm iso}$ distributions that are statistically similar to the observed $E_{\rm iso}$ distribution, with KS-test p-values at around 0.3 for both distributions (meaning we cannot reject the null hypothesis that these two distributions are drawn from the same parent population).  \\

\subsection{Isotropic Energy Evolution over Cosmic Time}
Motivated by recent results showing GRBs above a redshift of about $(1+z) \sim 2.5$ appear to be dominated by collapsar progenitors whereas lower redshift bursts may be dominated by non-collapsar (presumably compact object merger) systems \citep{LRBP26}, 
and expecting that these progenitor models may  have different jet structures, we separate our data into low and high redshift sub-samples at this redshift delimiter of $(1+z) = 2.5$, and ask the question: is there a difference in the best-fit jet structure between low and high redshift sub-samples?  \\

The answer appears to be that the general power-law behavior of the jet remains the same, but the normalization factor changes (with higher overall normalization for higher redshifts).  We show this in Figure~\ref{fig:lowzhiz}, where the orange histograms on the left show the observed and simulated higher redshift GRBs and the blue histograms on the right show the observed and simulated lower redshift GRBs.  The model used for each simulated distribution is shown by lines PL 3 and PL 4 in Table 1.  The change in energy normalization as a function of redshift in fact aligns with previous studies that have suggested that GRB isotropic equivalent energy and luminosity evolve with redshift \citep{LRFRR02,Yon04, KL06, Sal09, Sal12, Yu15, PKK15, TW15, Pesc16, Deng16, LR19c}.  \\

And indeed when we include the functional form of isotropic-equivalent energy evolution in our jet model, with the normalization evolving as $A \sim (1+z)^{k}$, where $1.4 < k < 2.0$ (the range of k values reported in the literature above), we find a fit to the overall distribution of isotropic-equivalent energy that is better than if no evolution is accounted for (as measured by a KS statistic). The left panel of Figure~\ref{fig:eisoevol} shows the observed and simulated distributions according to the model PL 5 in Table 1. A KS-test between the simulated and observed distributions in this case gives p-values $> 0.3$. Alternatively, within a top-hat jet picture, it has been shown that the observed isotropic energy evolution can in fact be explained by an evolution of the characteristic jet opening angle, rather than the magnitude of the energy in the jet \citep{LRFRR02, LR20, Khat25}.  We find that including this cosmic evolution in the characteristic jet angles ($\theta_{j}$ and $\theta_{m}$), rather than the energy normalization factor, does indeed reproduce the observed isotropic energy distribution equally well.  We show this in the right panel of Figure~\ref{fig:eisoevol}, described by model PL 6 in Table 1.\\

We also perform an additional exclusion analysis of the jet power-law index, scanning our parameter space to find the jet structures that {\em aren't} able to reproduce the $E_{\rm iso}$ distribution. This is shown in Figures~\ref{fig:zetafixedtheteff}, ~\ref{fig:zetafixedtheteffAcap}, and ~\ref{fig:zetafixedeff} in our Appendix section and described in section A of the Appendix.  What we can conclude from this additional analysis is the following:  {\em A power-law jet profile can indeed reproduce the spread in our energy distribution, but only for jet structures with power-law indices $\zeta \gtrsim -2.5$, for any reasonable set of priors and detector limits}.  Including redshift evolution of the normalization gives a tighter constraint on the jet profiles that better reproduce observations, still pointing to jet structures with power-law indices $\zeta \gtrsim 2.5$. We discuss what that may mean in terms of the physics of the jet in our Discussion section.

\subsection{Luminosity Distribution}
How does this jet structure do in reproducing the observed isotropic-equivalent peak luminosity distribution? Using the same power-law form for the jet structure from Model PL 1 in Table 1 (but with different normalization; we call this model PL A in Table 2) does not do well, overproducing high luminosity GRBs.  In fact, we find that the simulated isotropic-equivalent peak luminosity distribution generated from the model in Equation 6 cannot reproduce the data, {\em unless} we include luminosity evolution.  When we include luminosity evolution\footnote{Note that the observed cosmic luminosity evolution is stronger than the energy evolution and is tied to the fact that duration appears to evolve over redshift; see, e.g. \cite{LR19c}.}, then we can reproduce the observed distribution of peak luminosity quite well from our power-law jet model.  We show these results as Model PL B in Table 2 and Figure~\ref{fig:lum}.\\

We note that the best-fit jet structure for the luminosity distribution is slightly shallower than the structure preferred by the $E_{\rm iso}$ distribution.  Assuming that this is not related to unaccounted-for observational biases or selection effects, there are a couple of factors that may cause this. First, the luminosity we show here is the luminosity at the time that the flux peaks in the prompt emission, $L_{p} = dE/dt_{t, \rm peak}$.  The time of this peak power is expected to vary from burst to burst, depending on the details of the dissipation in the prompt phase, the emission mechanism itself, and the bulk Lorentz factor\footnote{And the bulk Lorentz factor will affect the timescale of the peak in the observer frame, with higher Lorentz factors leading to smaller peak timescales.} In other words, the angular structure fit for the peak luminosity is a characterization of an instantaneous emission efficiency at a time that may vary more widely from burst to burst, while the angular structure of isotropic energy reflects a time-integrated
distribution in the jet.  The slightly flatter structure may also simply indicate that emission in the wings is more efficient than the emission at the jet core.  This could be due to more efficient particle acceleration at the wings (e.g., the wings are more turbulent or less magnetized), similar to the limb brightening we see in radio images of AGN jets \citep{Nag14,Kim18,Hiro24, Park24}.

\begin{figure*}
\begin{centering}
    \includegraphics[width=0.45\textwidth]{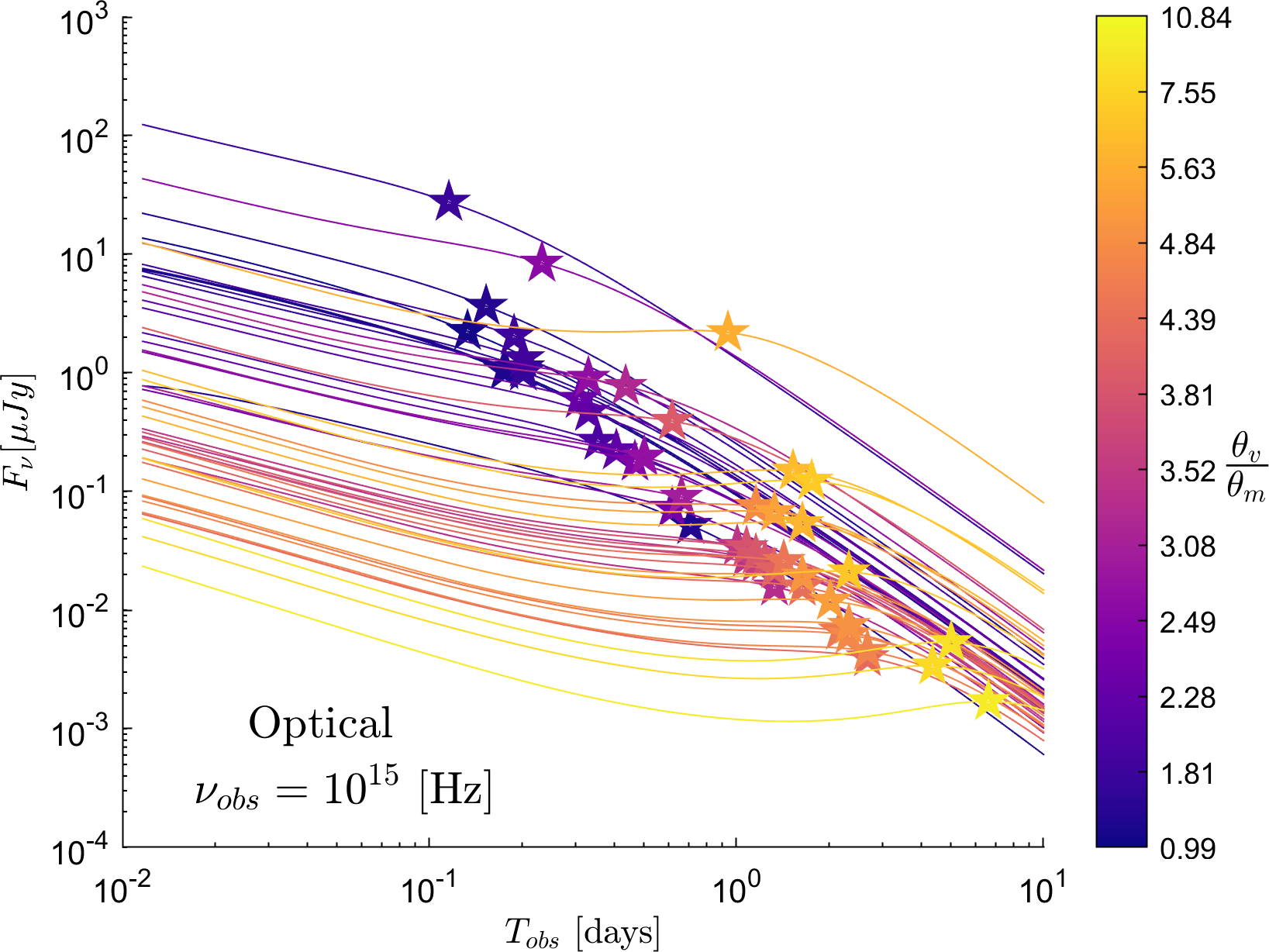}
     \includegraphics[width=0.45\textwidth, height=6.2cm]{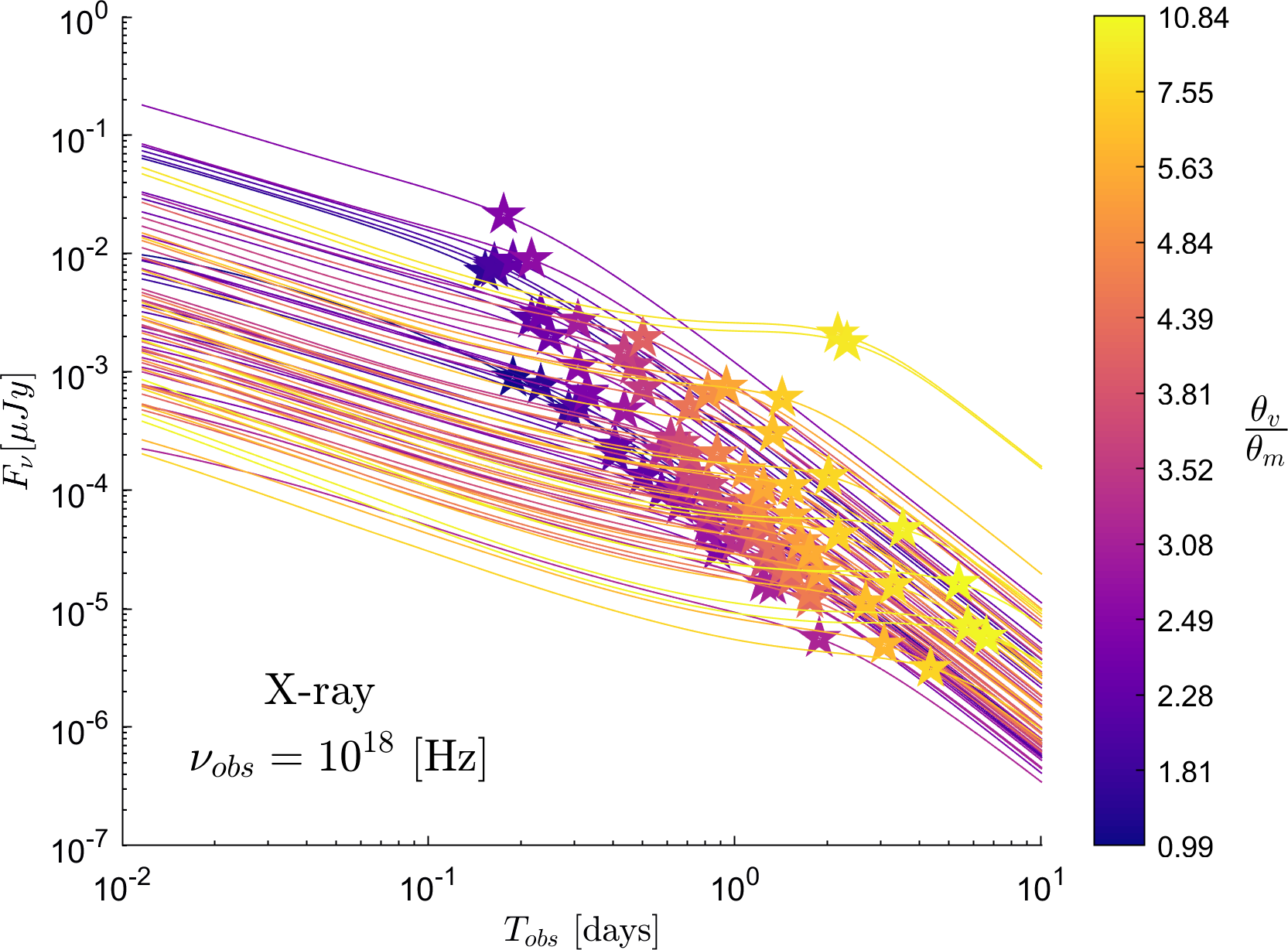}\\
    \caption{Afterglow light curves in the optical (left) and X-ray (right) bands predicted from our structured jet model PL 2, for fiducial microphysical parameters $p=2.2$, $\epsilon_{e},\epsilon_{B} = 0.01$, a bulk Lorentz factor of $\Gamma = 100$ and a single ISM density of $n_{\rm ISM} = 1 {\rm cm}^{-3}$.  Each light curve has a different redshift taken from the \cite{Zhao20} Table 1, and a viewing angle inferred from the listed $E_{iso}$ value in that table (see the detailed description in the text). The color bar shows this viewing angle normalized by the core jet angle $\theta_m$. The breaks are marked in stars.}
    \label{fig:lightcurves}
\end{centering}

\end{figure*}

\begin{figure*}
\begin{centering}
   {\includegraphics[width=0.48\textwidth]{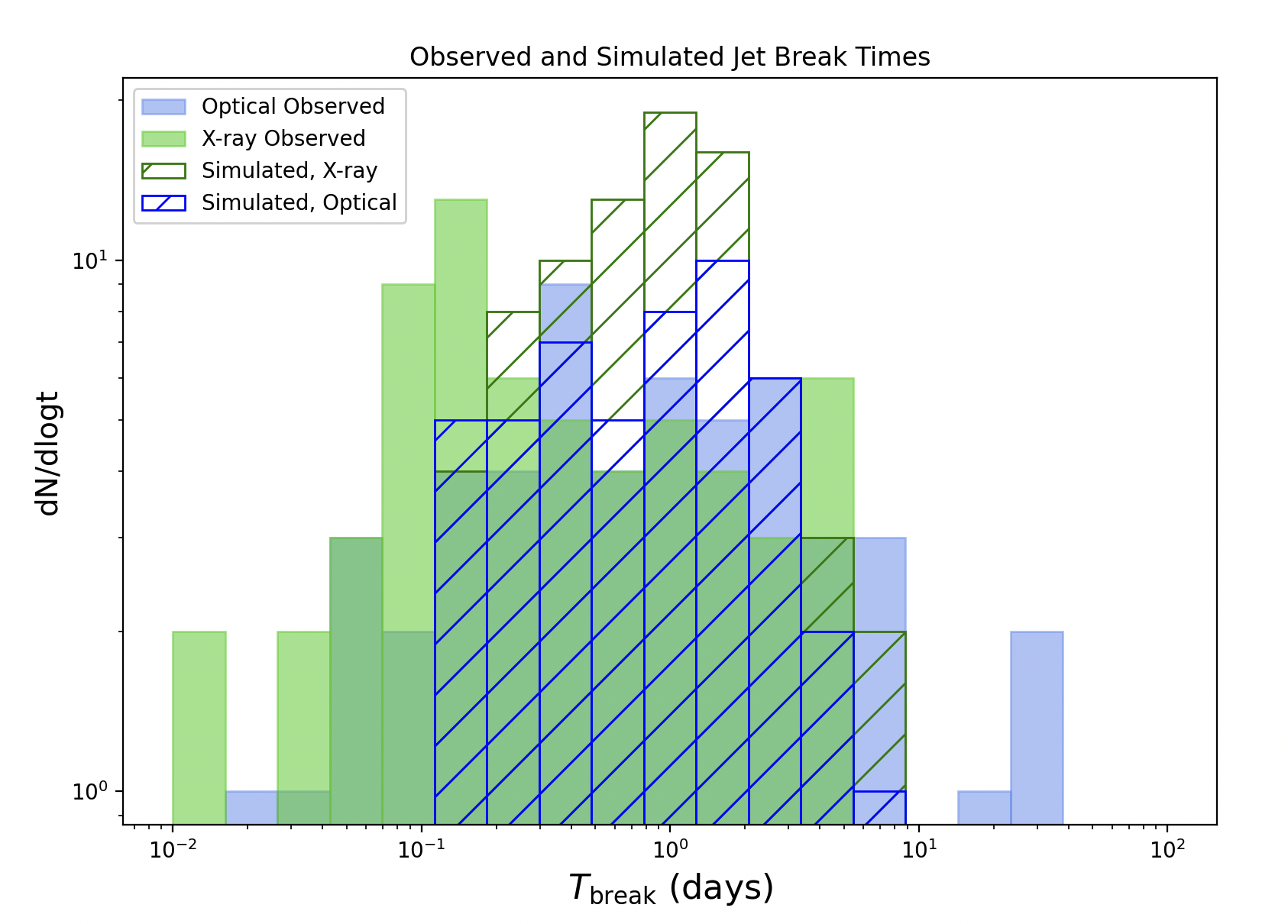}}
    \includegraphics[width=0.48\textwidth, height=6.2cm]{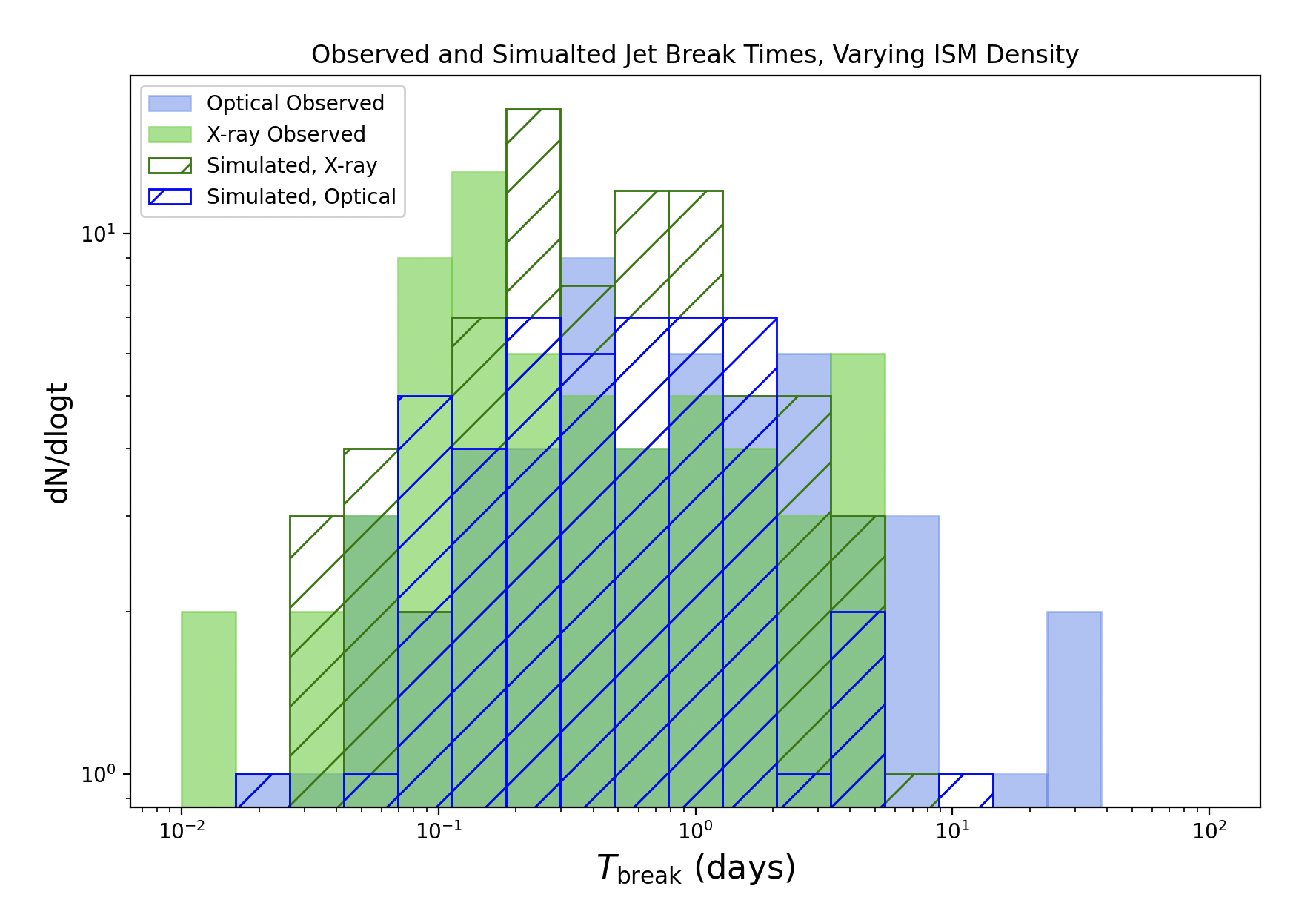}\\
    \caption{
    {\bf Left}: Simulated (hatched histograms) and observed (solid histograms) afterglow break times in the X-ray (green) and optical (blue) bands.  The simulated break times are calculated in the context of our structured jet model PL 2 in Table 1, as described in the text and in the previous figure caption (and shown by the stars in the previous figure).  
     {\bf Right}: Same as the right panel, but instead of using a single value for the ISM density for all bursts, we draw our ISM densities randomly between a range of $0.05 - 500 \ {\rm cm}^{-3}$. This naturally spreads out the distribution.}
    \label{fig:breaktimes}
\end{centering}

\end{figure*}

\section{Afterglow Light Curves and Jet break Times}

For the case of a structured jet, a break in the afterglow light curve reflects something slightly more subtle than in the top-hat jet model. The jet will still decelerate to a point such that the radiation is no longer relativistically beamed entirely along the observer's line of sight, when $\Gamma \sim 1/\theta_{v}$.  In this case, the beaming angle distribution is a rough reflection of observer viewing angle (as opposed to the physical jet opening angle as in the top-hat model).  \\

 To calculate the afterglow light curves and break time in our power-law jet model, we need our jet structure, a viewing angle, redshift, values for the microphysical parameters, a Lorentz factor, external density, and the wavelength band over which we would like the light curve generated. We then want to compare these predicted break times with observed GRB break times in the relevant energy band.  One way to do this is to simply use a uniform observer viewing angle distribution and redshifts sampled from the GRB redshift distribution (while making a specific choice for the GRB microphysical parameters, external density and Lorentz factor). \\

However, a more direct way to approach this is the following: \cite{Zhao20} provide a list of observed jet break times measured primarily in optical and X-ray bands (see their Table 1). For each GRB in the \citet{Zhao20} table, use their measured $E_{\rm iso}$ to calculate a corresponding viewing angle within our ``universal-jet" model (our inferred viewing angles are given in Figure~\ref{fig:thetav} in the Appendix).  Then -- given that viewing angle and given the observed redshift for that particular GRB, our given set of microphysical parameters, external density, and Lorentz factor -- we calculate the afterglow light curve in the respective bandpass.  In keeping with a theme of a universal jet, we choose a representative set of microphysical parameters for all of our light curves: $\epsilon_{B} = 10^{-2}$, $\epsilon_{e} = 10^{-2}$, $p = 2.2$. For simplicity, we choose a pre-deceleration Lorentz factor of $\Gamma_{o} = 100$ and constant density medium of $1 \ {\rm cm}^{-3}$ for each light curve, but discuss and show how our results change when we incorporate a more realistic distribution of densities and Lorentz factors.  \\

We show the predicted afterglow light curves from the power-law jet structure in Figure~\ref{fig:lightcurves}.
The light curves are generated according to the methods presented in \citet{GG18} and \citet{B24}, where the input jet structure is PL 2 in Table 1 and the rest of the afterglow model parameters are described above. The curves are colored according to their respective normalized viewing angles $\frac{\theta_{v}}{\theta_m}$, where darker colors correspond to the line of sight being closer to the edge of the jet core, while lighter colors represent jets viewed far from the core edge. We mark the jet break times with stars, which correspond to the geometrical break in both the optical and X-ray bands. This time is found by computing the slope of the light curve between every two points in time and marking the time in which the change in slope is maximal. Despite varying redshifts between the considered light curve models, the jet break time becomes more delayed as the value of the viewing angle grows and the system becomes more misaligned. \\

Figure~\ref{fig:breaktimes} shows both our simulated break times (hatched histograms) and the observed break times from the \cite{Zhao20} table (solid histograms).  The green colors indicate the X-ray band and the blue indicates the optical band. The left panel shows the break times for a single choice of circumburst density, $n=1$ cm$^{-3}$. In reality there is certainly a range of density profiles (of both ISM-like and wind-like) that will change the time of the break.  Given that the jet break time scales like $n^{-1/3}$ (for the ISM case), we expect our simulated break distribution to be spread by the range in density values to the 1/3 power.  The right panel of Figure~\ref{fig:breaktimes} shows the break times scaled by a range of densities pulled from a uniform distribution (in log space) between $0.05 -500$ ${\rm cm}^{-3}$.  As expected this spreads our break time distribution over a range more commensurate with the observational spread.  \\

 There are several important caveats that come into play here. On the observational side: 1) Many follow-up observations don't necessarily span the full time range necessary to accurately capture the jet break.  2) The observed break times themselves often have large error bars, and -- importantly -- the evolution of spectral frequencies in the bandpass may cause a steepening, leading to a false observed ``break time''.  On the theoretical side: 1) We chose a simple, single set of constant jet microphysical parameters ($\epsilon_{e}$, $\epsilon_{B}$, and $p$), which is certainly an oversimplification.  A more realistic spread in these parameters will spread out our simulated break time distributions. 2) We have also chosen a single Lorentz factor for all of the afterglow light curves.  As we discuss in the next section, this variable will have some amount of angular variation across the jet.  Accounting for this will broaden our break time distribution somewhat (but does not affect the time of the light curve break too much, as this is primarily determined by when we see the uniform core of the jet).  3) The simulated light curves may show a break due to spectral frequency evolution. However, we note that the core of the jet is the region that dominates the flux during the time of the geometrical light curve break. In the simulated optical light curves, the cooling frequency in this region can cross the observed bandpass at an observer time that depends on the redshift. But the light curve slope change for this kind of spectral crossing is relatively minor \citep{SPN98}, and is discernible from the geometrical break for an off-axis structured jet.  The simulated X-ray light curves (Fig. \ref{fig:lightcurves}, right panel) do not suffer from this effect, as the observing frequency remains well above $\nu_c$ in the jet core.  \\

\section{Discussion}
The concept of structured jets, and particularly quasi-universal structured jets, has been proposed since the time of the earliest GRB redshift confirmations and afterglow detections \citep[e.g.][]{Mesz98, DG01, Ross02, ZM02, KG03, Zhang04,LR04,DZ05}.
Similar to what we have done here, some previous studies have suggested a quasi-universal jet can explain broad population properties of GRBs, particularly the GRB luminosity function \citep[e.g][]{LR04,DZ05,Sal20}. For example, \cite{Sal20} claim the overall long duration GRB luminosity function is well-produced from a quasi-universal jet energy profile based on simulations by \cite{Laz17,LP19}.  Their jet energy profile has two distinct parts (corresponding to the jet itself and the surrounding cocoon), with an average structure that has a flat core until about 0.03 radians ($1.7^{\rm o}$), and an approximate power-law beyond this core, with an index of $\zeta \sim 3$ (see their Figure 3), similar to the structure we propose in this paper. \\

  There also exist a few specific GRBs that have allowed us to place observational constraints on their structure. For example,  GRB221009A (the so-called brightest of all time or ``BOAT'' GRB), GRB170817A (the gamma-ray burst associated with coincident gravitational wave emission from a neutron star merger), and GRB260310A (a nearby GRB associated with a broad-lined Type Ic supernova) have provided us unique opportunities to constrain jet structure because of their nearby distances and detailed broadband observations.  \cite{Ocon23} found that the broadband observations of the BOAT can be explained by a jet with a narrow core of about $3^{\rm o}$ \citep[although see the paper by][ who showed that there is degeneracy between determining the observer viewing angle and jet core opening angle from afterglow light curves alone]{NP21} and a power-law decline with an index of $1.15$, much shallower than what we find here (we show the predictions from such a shallow structure in our Appendix).   Meanwhile, GRB170817A was fit with a much steeper jet structure profile, with a power-law index between about 3 to 5, more commensurate with our results here.   Analysis of GRB260310A suggests \citep{Yang26} a bright jet core energy of about $10^{54}$ erg with a power-law decline as a function of angle, with an index of $\zeta \sim 2.5$. 

\subsection{Jet Structure from High Fidelity Simulations}
A number of studies have attempted to constrain GRB jet structure through simulations of jets in both collapsar/massive star and compact object merger ambient environments \citep{Laz17,Kat19,GNBMag,GNBHydro,Ur23,Ham23,Pav23,Pais24,Kat24,Ur25,Ur26}. The results vary, depending on how each simulation models the jet interaction with its surrounding medium, but most find a power-law like structure with indices between $1.5 \lesssim \zeta \lesssim 4$. For example, \cite{Kat19} calculated jet structure in a post-merger environment and found a power-law structure with an index of $\zeta \sim 3$.  When they included a dynamical ejecta component and embedded the system in an AGN disk \citep{Kat24}, they found the index was shallower, more around $\zeta \sim 2$.  \\

 \cite{GNBHydro} and \cite{GNBMag} performed 3D simulations of both hydrodynamic and weakly magnetized jets, respectively, propagating in both a massive star envelope and a compact object merger wind.  Their simulations showed that hydrodynamic jets have energy profiles that can be described as having a roughly constant energy core out to an angle of about 0.02 radians ($1.2^{\rm o}$) from the central axis, and then a power-law decline, with a power-law index $\zeta \lesssim 2$ out to jet angles of around 0.8 radians ($45^{\rm o}$) and then a steep decline thereafter. They argue the power-law index from their simulations depends strongly on the degree of mixing between the cocoon and the jet, with more mixing allowing the energy from the jet to be spread into the cocoon and create a shallower jet profile.  
 
   Meanwhile, their weakly magnetized jets, under the assumption of a specific stellar density profile and  toroidal magnetic field configuration, tend to show steeper power-law indices.  In this case, they find $\zeta \gtrsim 3$, because the magnetization tends to suppress the mixing and create a jet in which more of the energy is concentrated toward the center (i.e. a steeper jet). Our results fall more in line with this regime, suggesting indeed weak magnetization in these jets or a mixing prescription that has not yet been fully explored in the hydrodynamic simulation case.  \\

 \subsection{On the Angular Profile of the Lorentz Factor}
In our analysis, we have assumed that the Lorentz factor across the jet is high enough such that $1/\Gamma \ll \theta_{j}$, which ensures we are only receiving radiation from the viewing angle line of sight along the jet. Consider our jet energy profile models, where the energy is decreasing as a function of angle with a power-law index of $\zeta \sim 3$.  If we assume that the energy of the jet (when it begins to radiate prompt emission) is mostly stored in kinetic energy,
\begin{equation}
    E_{K} \propto \Gamma^{2} n m_{p} c^{2}
\end{equation}
and that this external density $n$ is relatively constant as a function of angle, this would require that the Lorentz factor have a power-law decrease as a function of angle with an index $\zeta/2 \sim 1.5$.  For example, a $\Gamma$ of 1000 at the jet core decreases to a $\Gamma$ of $\sim 10$ at our jet wings, still satisfying our condition of $1/\Gamma < \theta_{j}$. As we mentioned in \S 3, a relatively flat Lorentz factor profile across the jet is also supported by both simulations of \cite{Kat19, BN19, Kat24}.   Alternatively, if the Lorentz factor profile were constant, we would need the mass density decreasing as a function of angle (at the time the jet radiates) or the time-integrated radiative efficiency changing as a function of angle (see the discussion in \S 3.2 regarding the luminosity profile). \\

\section{Summary and Conclusions}
In this paper, we have asked the question of whether a single ``quasi-universal'' gamma-ray burst jet angular structure (energy as a function of angle from the jet axis) can reproduce the observed isotropic energy and luminosity distributions, where the observer viewing angle is the only variable that determines the spread in the distributions. {\bf We find a power-law angular structure consistent with weakly magnetized jets can indeed reproduce the observations, particularly when the overall energy normalization evolves with cosmic redshift}.  Such a jet can also accommodate the distribution of observed break times in afterglow light curves. \\

Our results can be summarized as follows:\\

\begin{itemize}
    \item The relatively broad isotropic equivalent energy distribution can be reproduced by a quasi-universal GRB jet with a power-law dependence of jet energy with angle from the central jet axis, where the observer viewing angle is what determines the spread in the energy distribution.  The power-law index is $3 \lesssim \zeta \lesssim 4$, consistent with structure predicted by simulations of weakly magnetized jets, but steeper than what is predicted for the structure of purely hydrodynamic jets.

    \item Including evolution of either the jet normalization energy over cosmic redshift ($A \propto (1+z)^{1.5\pm 0.5}$) or the characteristic jet angles ($\theta_{j,m} \propto (1+z)^{-1.0\pm 0.2}$) provides an even better match to the data, in terms of a KS test comparison between the predicted and observed $E_{\rm iso}$ distributions.  The functional forms of the cosmological evolution align with those of previously published studies.

     \item The peak luminosity distribution can be produced by a power-law angular structure, only if we include its dependence on cosmic redshift according to a functional form published in previous studies.   The peak luminosity distribution can accommodate shallower power-law indices, but still in the realm of a weakly magnetized jet $\zeta \gtrsim 2.5$.

    \item The predicted distribution of afterglow break times from this quasi-universal jet overlaps well with the observed distribution of afterglow jet break times, although it is slightly narrower.  This is true even for a single choice of microphysical parameters and external circumburst density. A more realistic treatment -- e.g. allowing for variation in the external density profiles across GRBs -- broadens the predicted jet break time distribution and reproduces the spread seen in the observations.\\

\end{itemize}

We have chosen one particular parameterization of the jet angular energy distribution and shown this can describe the set of observations we consider in this paper.  Of course, the observed $E_{\rm iso}$ distribution could be explained by truly different intrinsic jet energies within a top-hat jet structure. Furthermore, more complicated non-monotonic jet structures (e.g. a power-law with an upturn or other features that may reflect interactions at, for example, the jet-cocoon interface) with enough tunable parameters may equally reproduce the observed distributions we present here. The point of this paper is merely to show how a single, remarkably simple jet structure - motivated by both high fidelity simulations and observations - is able to reproduce the wide spread in observed or inferred GRB properties.  That a quasi-universal structure can achieve this may indeed reflect the fundamental physics governing relativistic jets and their interaction with their cocoons/ambient surroundings. \\

Future observations of GRBs in new parameter spaces -- for example, fast X-ray transients that have been proposed as off-axis GRB jets \citep{Chen26} and polarization measurements that can offer diagnostics of jet structure \citep{Lask26} -- may better directly constrain the angular structure of the jet and allow us to ultimately pin down the physics of GRB jets.

\begin{acknowledgments}
We thank James Leung for interesting and thoughtful discussions on GRB observed distributions and afterglow physics.  N.L.-R. thanks Isaac Diaz-Ray for helpful discussions on statistical approaches to this problem, Adithan Kathirgamaraju for interesting discussions and references on GRB jet structure from simulations, and Tsvi Piran for elucidating conversations on GRBs in general.  FDC acknowledges support from the DGAPA/PAPIIT grant IN113424. GB is supported by the President's Fellowship for excelling postdoctoral researchers in the exact sciences from the Open University of Israel and grant no. 1649/23 from the Israel Science Foundation.  This work was supported by the U.S. Department of Energy through Los Alamos National Laboratory (LANL).  LANL is operated by Triad National Security, LLC, for the National Nuclear Security Administration of U.S. Department of Energy (Contract No. 89233218CNA000001), LA-UR-26-27059.\\
\end{acknowledgments}

\begin{contribution}

All authors contributed equally to the collaboration.

\end{contribution}


\bibliography{refs}{}
\bibliographystyle{aasjournalv7}

\appendix

\section{Additional Jet Structures - Exclusion Analysis}
\begin{figure}
\begin{centering}
    \includegraphics[width=0.45\textwidth]{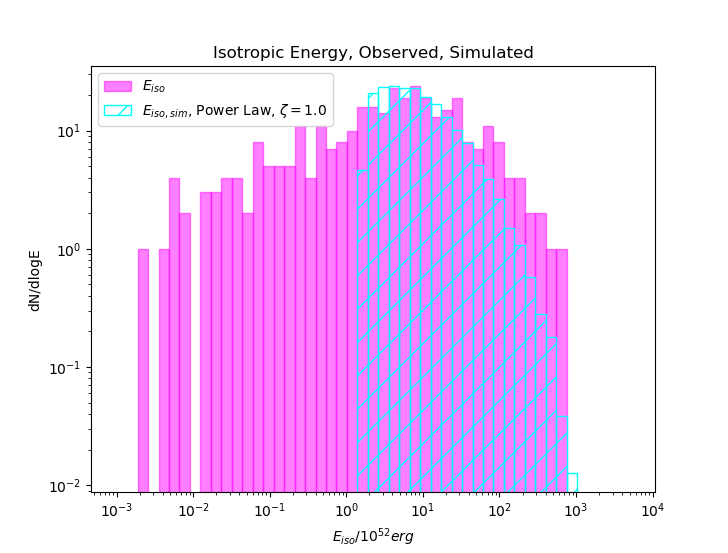}\includegraphics[width=0.45\textwidth]{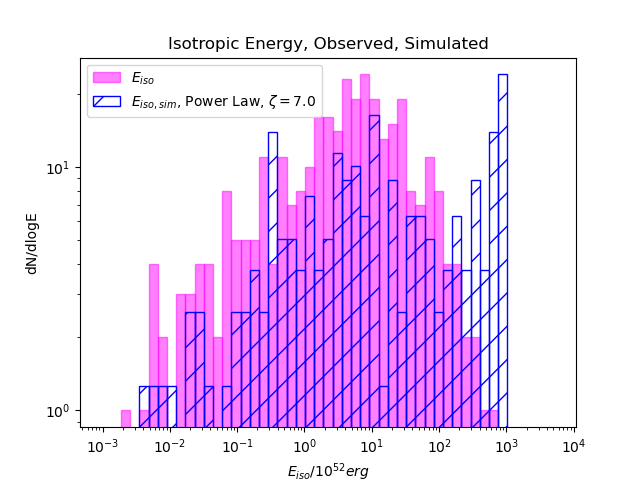}
    \caption{The isotropic energy distribution produced from a power-law jet profile (blue hashed histograms) compared to the observed distribution (magenta), for an energy profile with power law index $\zeta = 1$ (left) and $\zeta=7$ (right). The shallow structure does not allow for low energy GRBs while the steep structure overproduces high energy GRBs.}
    \label{fig:extgrbjet}
\end{centering}
\end{figure}

There is an important interplay between the parameters in our model described by equation 6, as well as the radiative efficiency, the redshift, and the detector limit.  For example, low redshifts, high radiative efficiencies, or setting too low of a detector fluence limit will all overproduce the low energy end of the $E_{\rm iso}$ distribution. Too high of a $\theta_{m}$ combined with a high normalization factor $A$ will overproduce the high end of the isotropic-equivalent energy distribution.  And of course the power-law index plays a key role in the shape of the $E_{\rm iso}$ distribution: As representative examples, Figure~\ref{fig:extgrbjet} shows the predicted $E_{\rm iso}$ distributions for jets with very shallow indices (left panel, $\zeta = 1$) and very steep indices (right panel, $\zeta = 7)$.  The shallow structure does not capture the low energy end of the distribution, while the very steep structure overproduces high $E_{\rm iso}$ GRBs. \\ 

 To explore how these degeneracies might affect the robustness of our results, we perform an exclusion analysis, searching the range of power-law indices that {\em cannot} reproduce the observed distributions, regardless of how we allow other parameters in the model to vary.  An important condition when we simulate an $E_{\rm iso}$ distribution is to ensure we have enough ``surviving'' GRBs that make it past the detector flux or fluence threshold in order to do a reasonable statistical comparison with the observed distribution.  We set this threshold at $N_{\rm survive}=300$. \\
 
 Following the same methodology as described in Section 3 for generating our $E_{\rm iso}$ distribution, we scan over the parameter space and find that the isotropic energy distribution strongly prefers jet structure power-law indices $ 2.5 \lesssim \zeta \lesssim 5$. If we impose additional constraints, like setting an upper limit on the energy normalization or fixing the core angle of the jet, the constraint becomes tighter with $ 3 \lesssim \zeta \lesssim 4$. 
We note that if we allow the radiative efficiency to be another free parameter, the $E_{\rm iso}$ indices are less constrained, allowing for steeper jet profiles (although flatter profiles are still excluded).  \\

\begin{figure}[h]
\begin{centering}
    \includegraphics[width=0.45\textwidth]{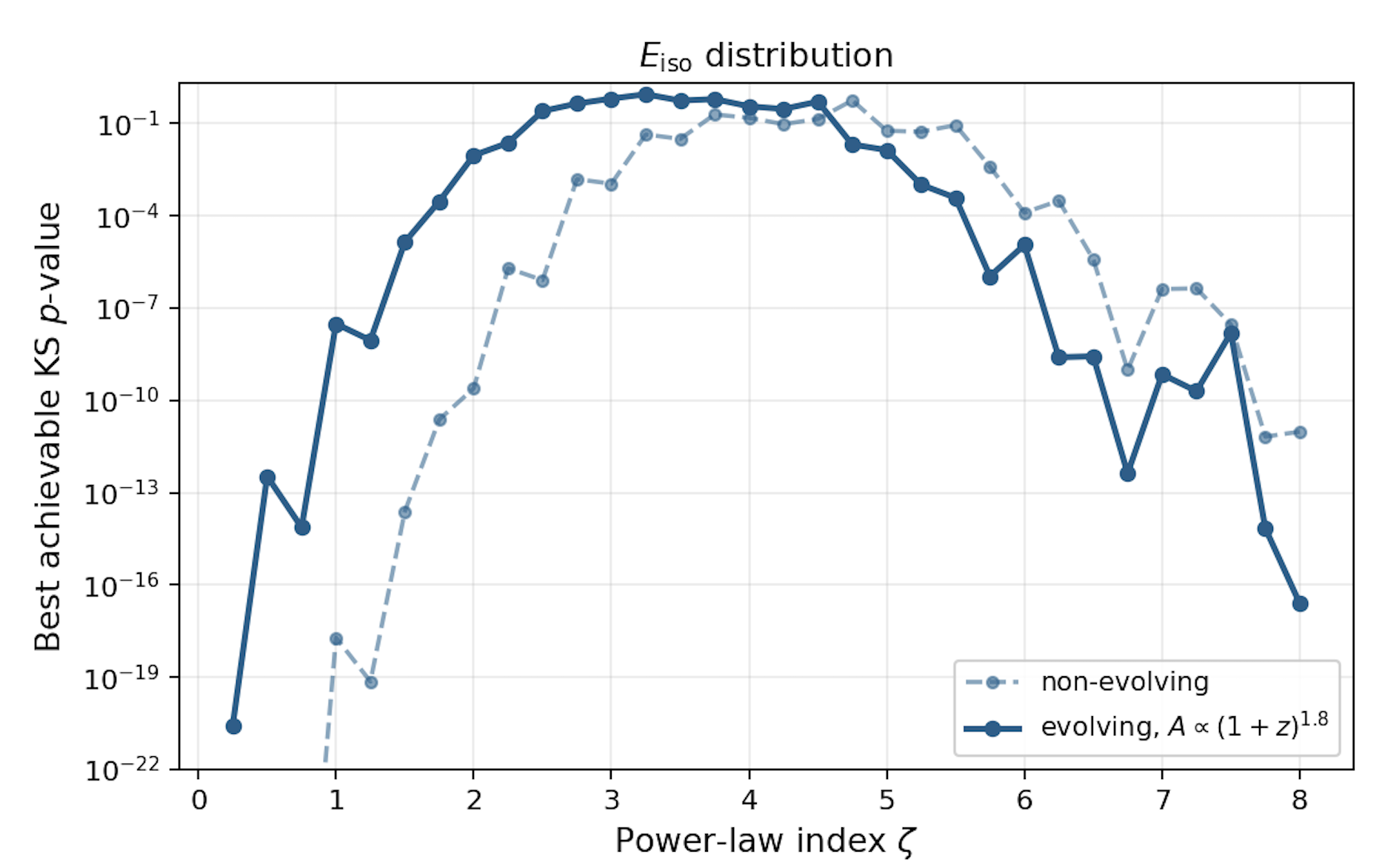}
    \includegraphics[width=0.45\textwidth, height=5.2cm]{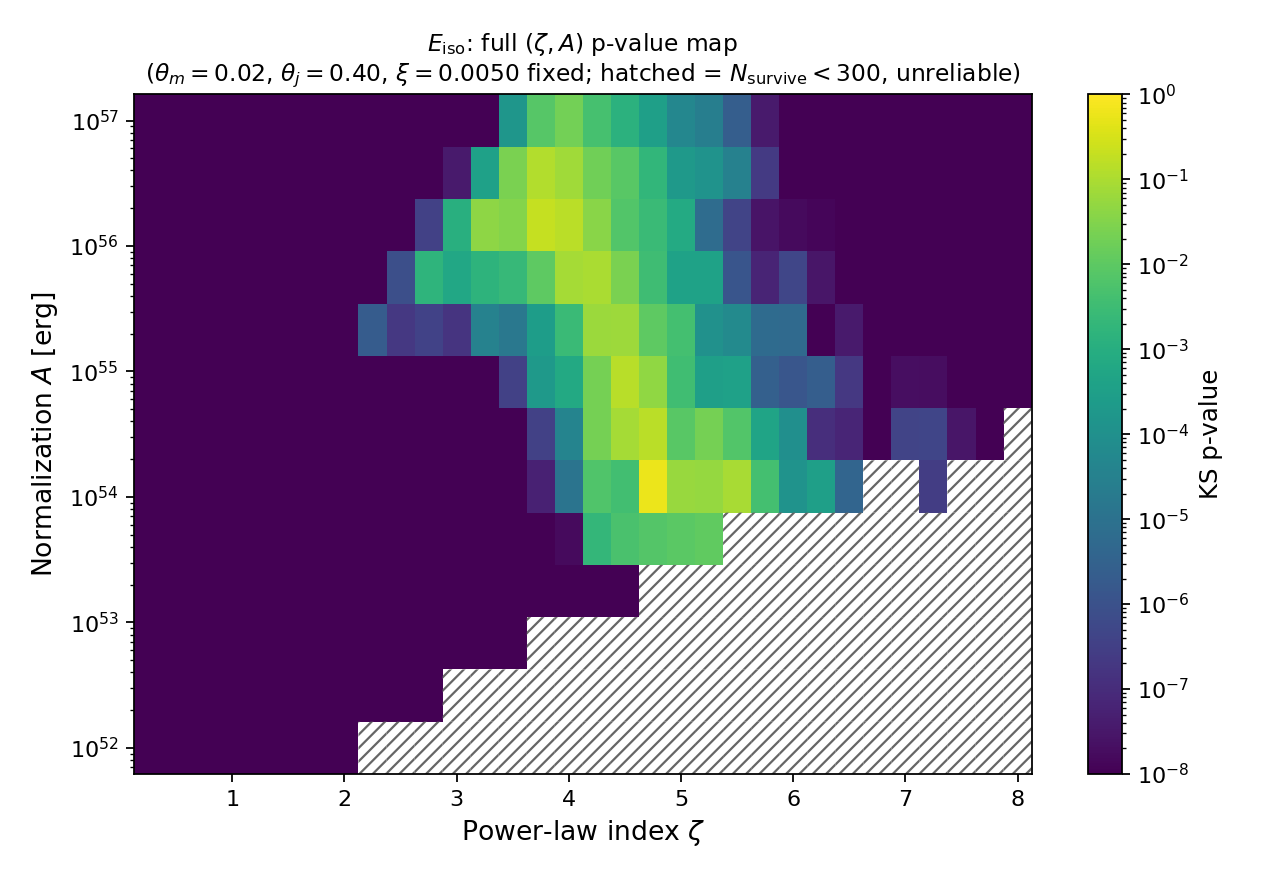}
    \caption{KS p-values between simulated and observed distributions as a function of jet structure power-law index.  The left plot (blue lines) shows the $E_{\rm iso}$ distribution comparison with (solid lines) and without (dotted lines) redshift evolution of the normalization energy (see \S 3.1 in text).  The right plot shows the KS values in the $A-\zeta$ plane, with the hatched region showing the parameter space where the surviving sample size was too small to make a fair statistical comparison. This parameter scan fixed radiative efficiency $\xi = 0.005$,  $\theta_{m} = 0.02$,  $\theta_{j}=0.4$ allowing only $A, \zeta$ to vary.}
    \label{fig:zetafixedtheteff}
\end{centering}
\end{figure}

\begin{figure}[h]
\begin{centering}
    \includegraphics[width=0.45\textwidth]{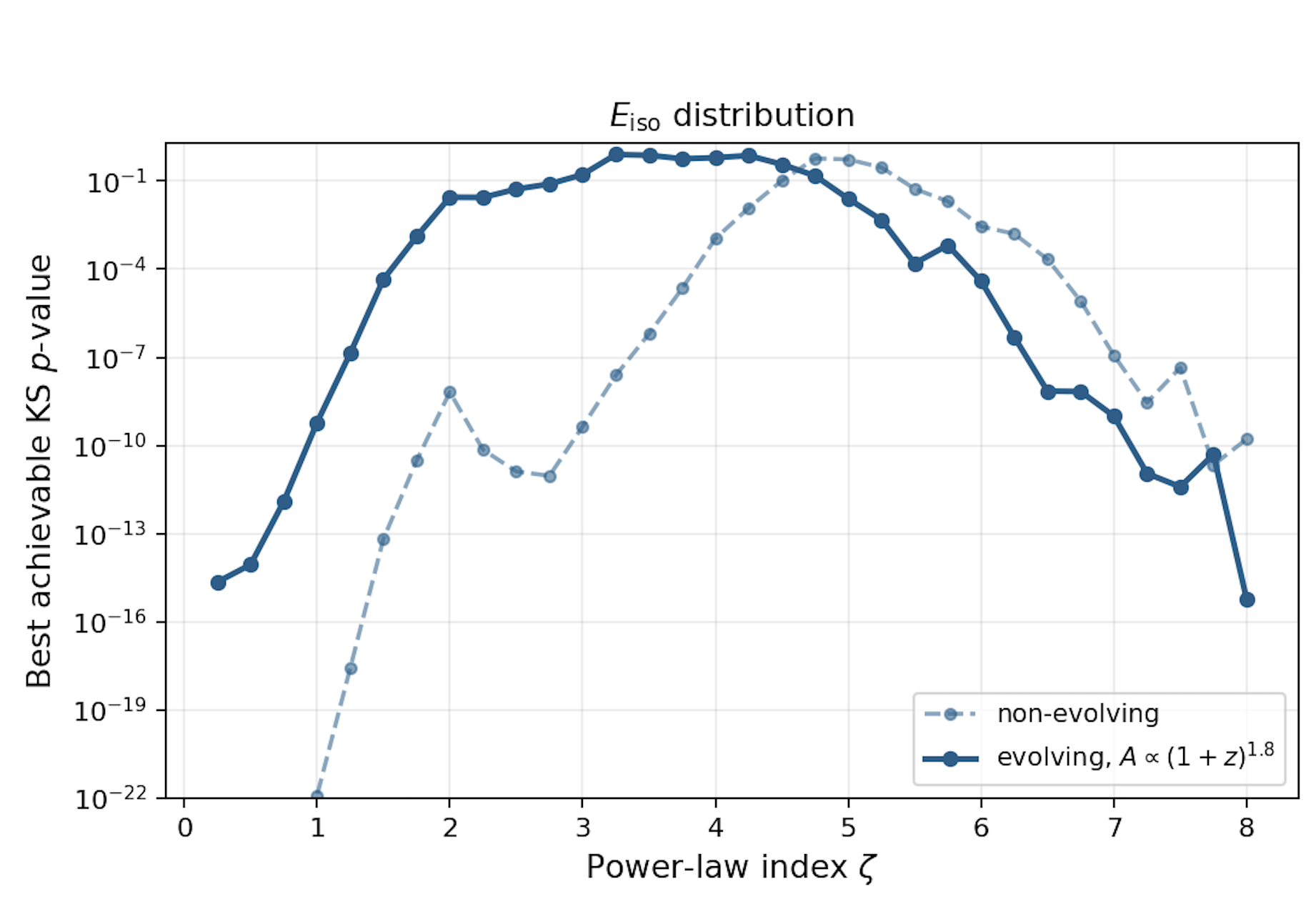}
    \includegraphics[width=0.45\textwidth, height=5.2cm]{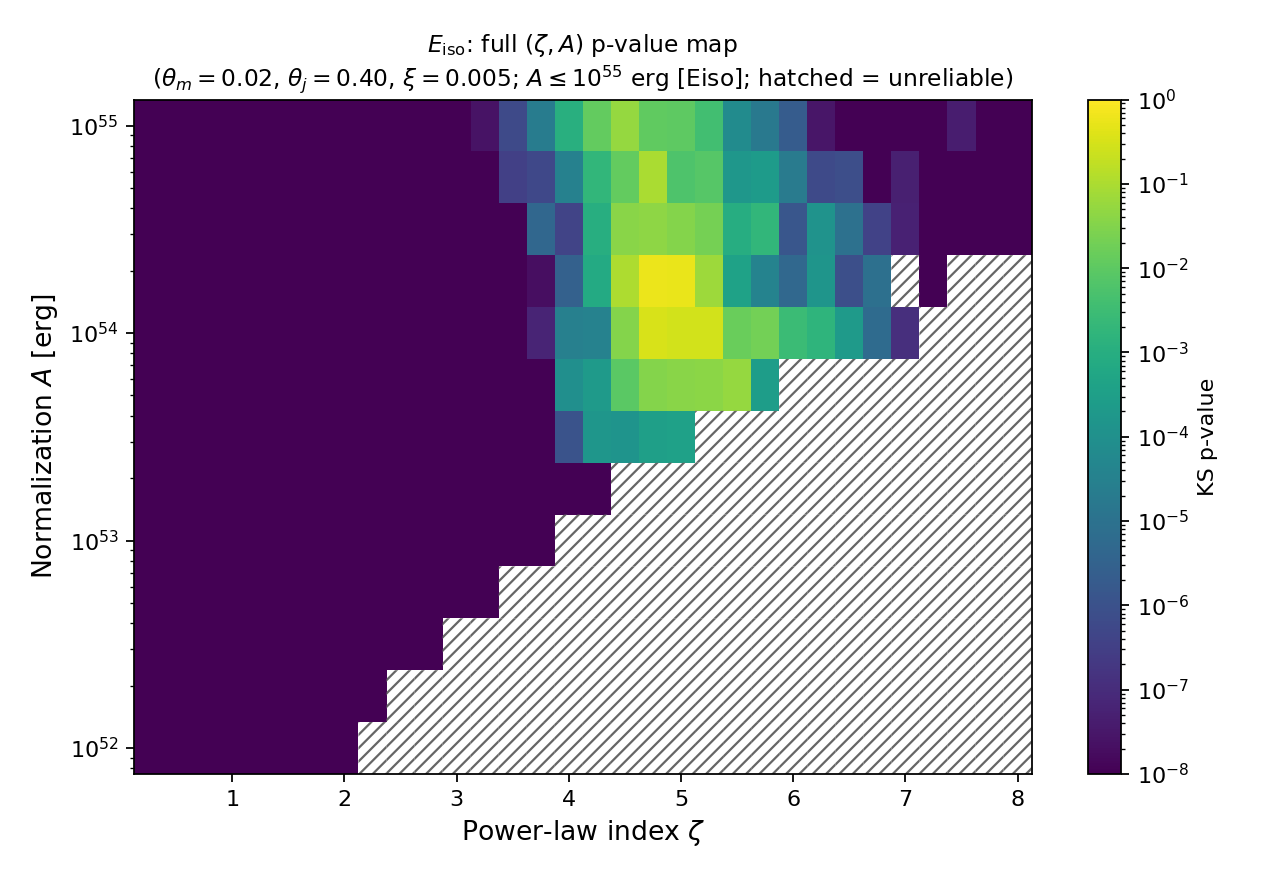}
    \caption{KS p-values between simulated and observed distributions as a function of jet structure power-law index.  The left plot (blue lines) shows the $E_{\rm iso}$ distribution comparison with (solid lines) and without (dotted lines) redshift evolution of the normalization energy (see \S 3.1 in text).  The right plot shows the KS values in the $A-\zeta$ plane. This parameter scan fixed radiative efficiency $\xi = 0.005$,  $\theta_{m} = 0.02$,  $\theta_{j}=0.4$ allowing only $A, \zeta$ to vary. In this case, however, $A$ was capped at a maximum value of $10^{55}$ erg.}
    \label{fig:zetafixedtheteffAcap}
\end{centering}
\end{figure}

\begin{figure}[h]
\begin{centering}
    \includegraphics[width=0.45\textwidth]{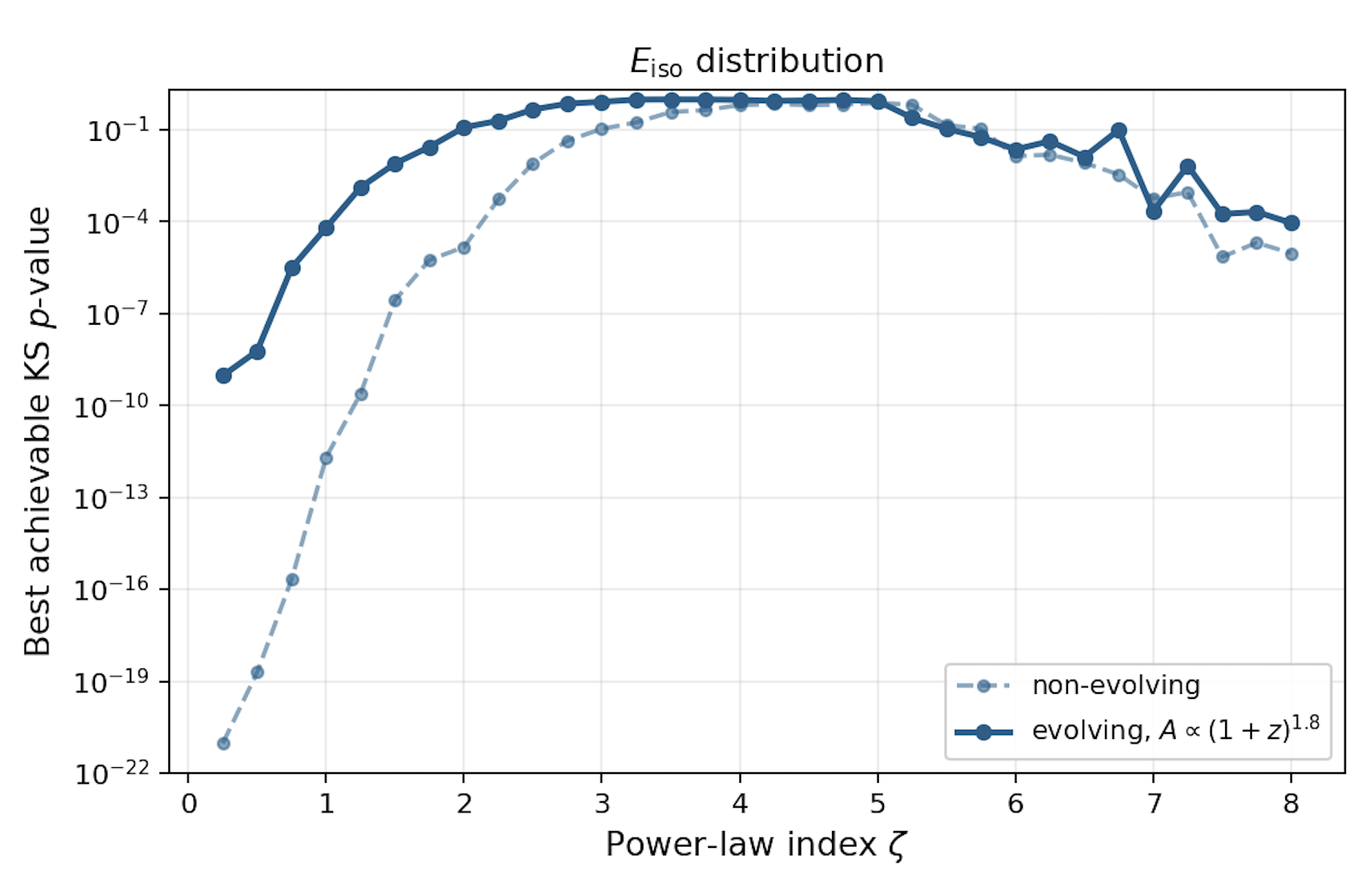}
    \includegraphics[width=0.45\textwidth, height=5.2cm]{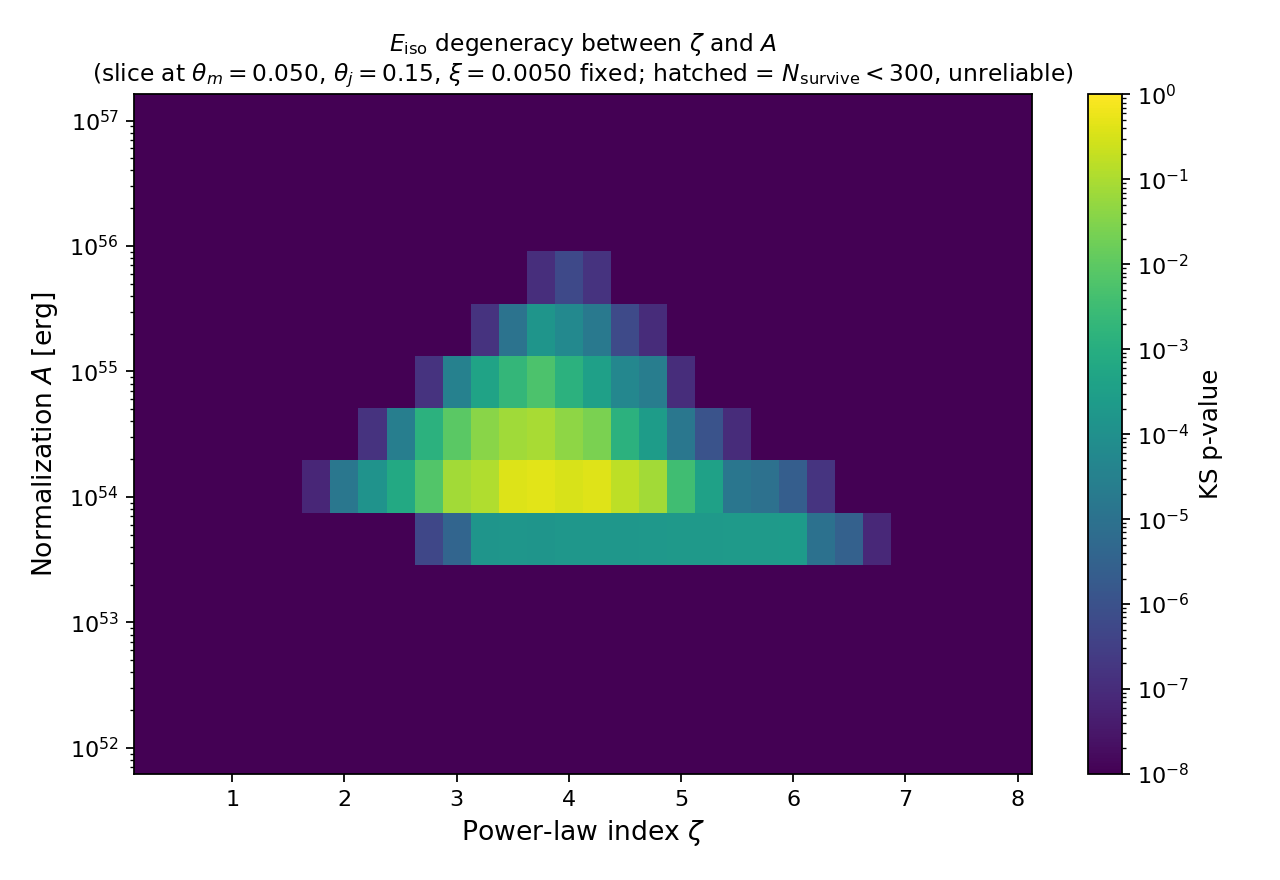}
    \caption{KS p-values between simulated and observed distributions as a function of jet structure power-law index.  The left plot (blue lines) shows the $E_{\rm iso}$ distribution comparison with (solid lines) and without (dotted lines) redshift evolution of the normalization energy (see \S 3.1 in text).  The right plot shows the KS values in the $A-\zeta$ plane.  This parameter scan allowed $\theta_{m}, \theta_{j}, A, \zeta$ all to vary, while keeping only the radiative efficiency fixed.}
    \label{fig:zetafixedeff}
\end{centering}
\end{figure}

Figures ~\ref{fig:zetafixedtheteff},  ~\ref{fig:zetafixedtheteffAcap}, and ~\ref{fig:zetafixedeff} show the KS p-values between simulated and observed distributions as a function of jet structure power-law index. We consider any p-value above about 0.1 to be sufficiently high that we cannot reject the null hypothesis that the distributions come from distinct parent populations. The left plots (blue lines) show the $E_{\rm iso}$ distribution comparison results, where solid lines include redshift evolution of the normalization factor $A$, but dotted lines do not (see \S 3.1 in text).  The right plots show the KS values in the $A-\zeta$ plane, with the hatched region showing the parameter space where the surviving sample size was too small to make a fair statistical comparison between the observed and simulated samples. \\

\section{GRB Redshift Distribution}
\begin{figure}
\begin{centering}
\includegraphics[width=0.6\textwidth]{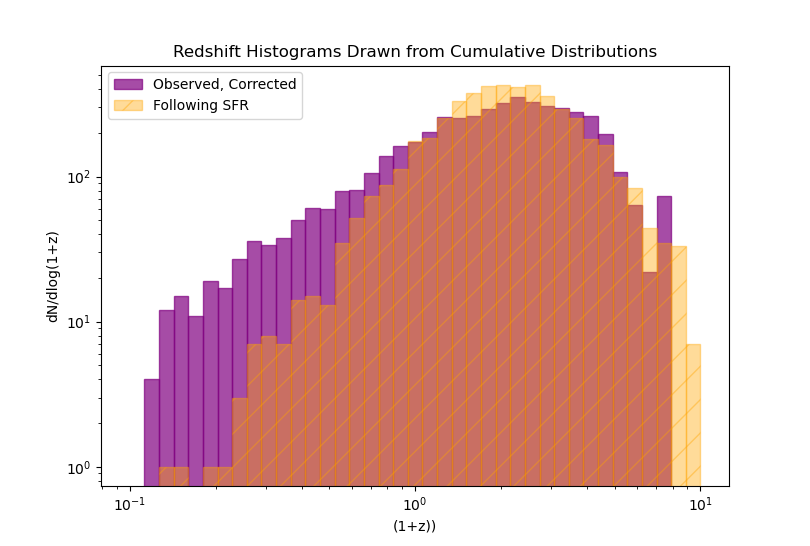}
    \caption{Differential distribution of GRB redshifts. The purple represents the observed GRB redshift distribution corrected for Malmquist bias, while the orange histogram represents a redshift distribution that follows the Madau-Dickinson Star Formation Rate \citep{MD14}.}
    \label{fig:grbsfrz}
\end{centering}
\end{figure}

\begin{figure}
\begin{centering}
    \includegraphics[width=0.6\textwidth]{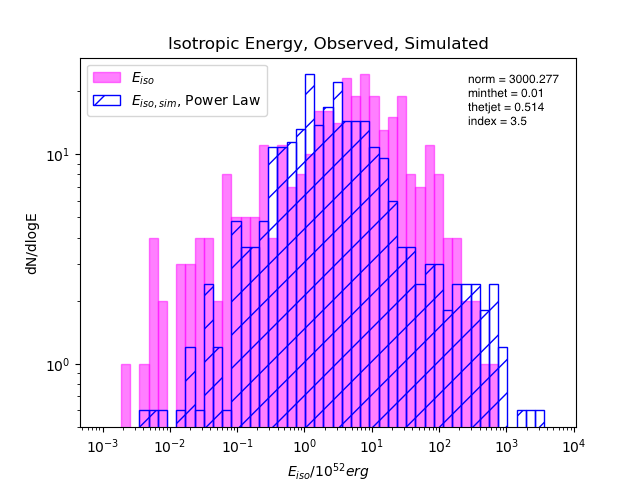}
    \caption{ The isotropic energy distribution produced from a power-law jet profile (blue hashed histogram) compared to the observed distribution (magenta), {\em assuming that the GRB redshift distributions directly follows the Madau-Dickinson star formation rate}. The best model has $\theta_{m} = 0.021 \pm 0.01$, an exponential cutoff at $\theta_{j} = 0.514 \pm 0.23 $, and a power-law index of  $\zeta = 3.53 \pm 0.50$. Because of the lack of dispersion in this redshift distribution (particularly at low redshifts; see Figure 10), this model struggles to capture the full width of the energy distribution, under-producing at the low energy end.}
    \label{fig:Eisosfr}
\end{centering}
\end{figure}

In our methodology in the main paper, we assigned our simulated GRBs a redshift according to the GRB redshift distribution corrected for sample incompleteness/Malmquist bias \citep{LRFRR02, LR19c}.  Figure~\ref{fig:grbsfrz} shows the GRB redshift distribution corrected for this bias as well as a redshift distribution that follows the star formation rate of \cite{MD14} exactly (hereafter MD14).  Figure~\ref{fig:Eisosfr} shows the resulting predicted $E_{\rm iso}$ distribution for GRBs directly tracing the SFR, and for a jet structure with $\theta_{m} = 0.021 \pm 0.01$, an exponential cutoff at $\theta_{j} = 0.514 \pm 0.23 $, and a power-law index of  $\zeta = 3.53 \pm 0.50$.   The universal jet structure fails to capture the full width of the $E_{\rm iso}$ distribution in this case.  

\section{Inferred Viewing Angles}
\begin{figure}
\begin{centering}
\includegraphics[width=0.7\textwidth]{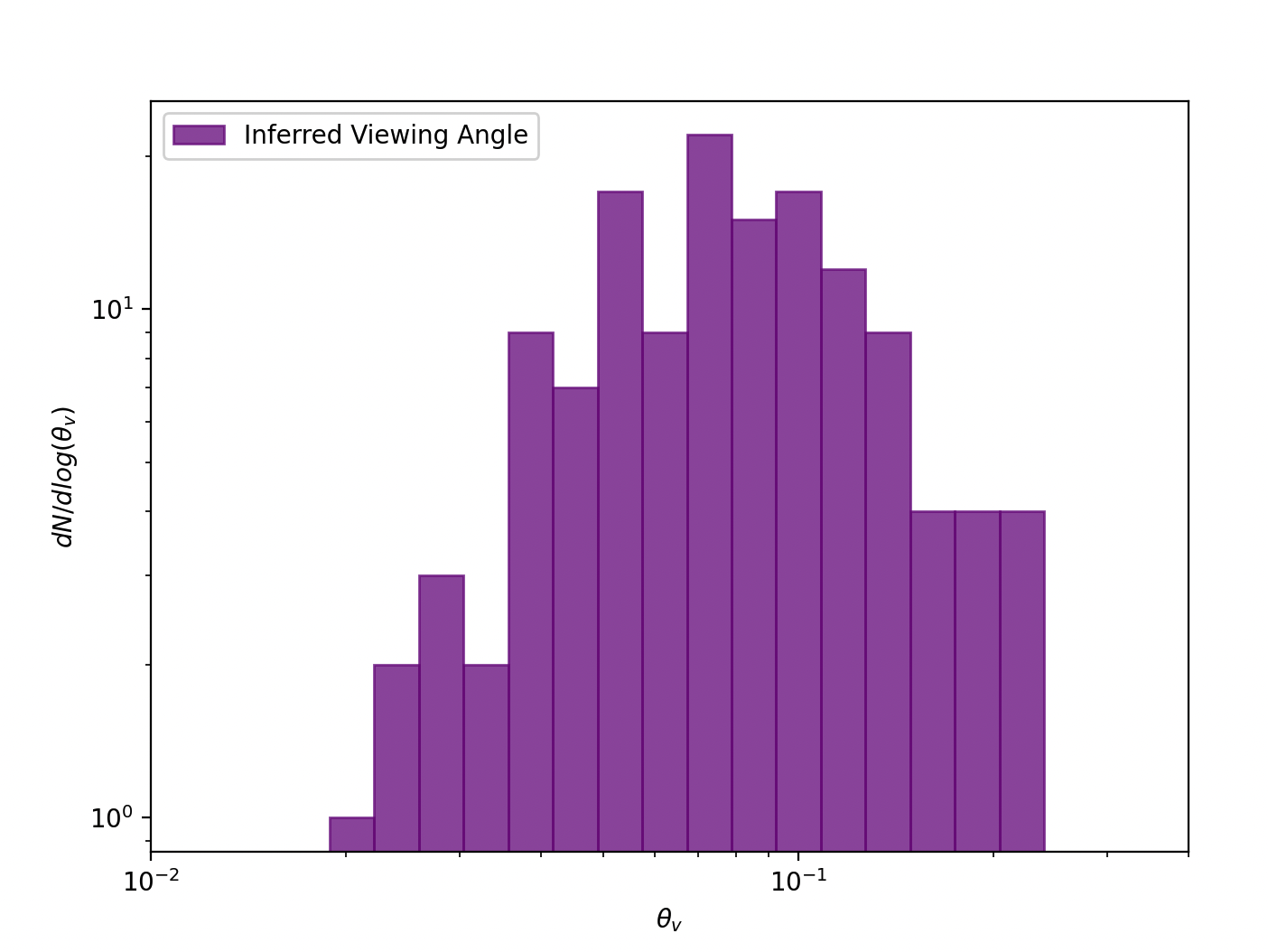}
\caption{Inferred viewing angles for GRBs presented in \citet{Zhao20}.  We used their inferred $E_{\rm iso}$ value to calculate the viewing angle according to our model PL 2 given in Table 1.}
\label{fig:thetav}
\end{centering}
\end{figure}

Finally, in Figure~\ref{fig:thetav}, we show the distribution of viewing angles we use in our afterglow light curve and jet break time calculations. This is done, as described in \S 4, by using our equation 6 with model PL 2 to calculate the viewing angle from the given value of $E_{\rm iso}$ in the \cite{Zhao20} paper.

\end{document}